\documentclass[journal]{IEEEtran}
\usepackage[pdftex]{graphicx}
\usepackage{algpseudocode}
\usepackage{amsmath,amssymb} 
\usepackage{xcolor}
\newcommand{\COMMENT}[1]{$\triangleright$  #1}
\newcommand{\C}[1]{$\triangleright$  #1}

\usepackage{comment}
\usepackage{hyperref}
\usepackage{textcomp}  

\usepackage{pgfplots}
\usepackage{etoolbox}
\usepgfplotslibrary{dateplot}
\pgfplotsset{compat=1.12}
\pgfplotsset{every axis/.append style={
		tick label style={font=\footnotesize}  
}}

\graphicspath{{./figures/}} 
\DeclareGraphicsExtensions{.pdf} 

\usepackage{booktabs}
\usepackage{multirow}

\usepackage{tikz}
\usetikzlibrary{shapes.geometric, arrows, positioning, calc, fit, backgrounds} 

\usepackage[indention=10pt, singlelinecheck=false]{subfig}

\usepackage{cite}  

\usepackage{wrapfig}

\usepackage{comment}

\tikzset{
	|-/.style={   
		to path={ (\tikztostart) |- (\tikztotarget) \tikztonodes }
	},
	-|-/.style={  
		to path={
			(\tikztostart) -| ($(\tikztostart)!#1!(\tikztotarget)$) |- (\tikztotarget)
			\tikztonodes
		}
	},
	-|-/.default=0.5, 
	|-|/.style={   
		to path={
			(\tikztostart) |- ($(\tikztostart)!#1!(\tikztotarget)$) -| (\tikztotarget)
			\tikztonodes
		}
	},
	|-|/.default=0.5,  
}

\tikzstyle{DB}     = [cylinder, line width=0.5mm, text width=2cm, text centered, 
                      shape border rotate=90, aspect=0.25, draw]

\tikzstyle{EMPTY} = [draw=none, text width=2.5cm, text centered, font=\fontsize{15}{0}\selectfont] 

\tikzstyle{BOX}   = [draw=red,         
                     ultra thick,   
                     text centered,
                     rectangle,        
                     text width=2.5cm] 

\tikzstyle{BOXL}   = [draw=red,         
                     line width=3pt,   
                     text centered,
                     rectangle,        
                     text width=2.5cm] 
                     
\tikzstyle{BOXL2}   = [	draw=red,         
						line width=2pt,   
						text centered,
						rectangle,        
						text width=3cm,
						font=\fontsize{15}{0}\selectfont] 
						
\tikzstyle{BOXL3}   = [	draw=red,         
line width=1pt,   
text centered,
rectangle,        
text width=3cm,
font=\fontsize{15}{0}\selectfont] 

\tikzstyle{BOXLD}   = [	draw=red,         
						line width=3pt,   
						dashed,			  
						text centered,
						rectangle,        
						text width=2.5cm] 
						
\tikzstyle{BOXLD2}   = [	draw=red,         
line width=2pt,   
dashed,			  
text centered,
rectangle,        
text width=2.5cm,
font=\fontsize{30}{0}\selectfont] 

 \tikzstyle{BOXB}   = [draw=black,         
					 line width=3pt,   
					 text centered,
					 rectangle,        
					 text width=2.5cm] 

\tikzstyle{BOXs} = [draw=red,         
                     ultra thick,   
                     text centered,
                     rectangle,        
                     text width=2cm] 

\tikzstyle{BOXC}  = [draw=red,       
                     ultra thick, 
                     text centered,  
                     rectangle,      
                     text width=3cm] 

\tikzstyle{queryComm}   = [fill=yellow!20,rounded corners, draw=black!50, dashed]
\tikzstyle{designComm} = [fill=blue!20,rounded corners, draw=black!50, dashed]

\tikzstyle{ARROW}  = [ultra thick, ->, >=stealth]             
\tikzstyle{DARROW} = [dashed, ultra thick, ->, >=stealth]     
\tikzstyle{IARROW} = [draw=blue, fill=blue, ultra thick, ->]  
\tikzstyle{OARROW} = [draw=red, ultra thick, ->] 

\tikzstyle{SBOX}   = [draw=blue, 
              ultra thick,       
              text centered,
              rectangle,         
              text width=1.6cm]  

\begin{document}

%
\title{Image Processing Methods for Coronal Hole Segmentation, Matching, 
	and Map Classification}
%
%
%
\author{V.~Jatla,~\IEEEmembership{Member,~IEEE,}
	M.S.~Pattichis,~\IEEEmembership{Senior Member,~IEEE,}      
	and C.N.~Arge
	\thanks{V.~Jatla and M.S.~Patthics are with the Department of 
		Electrical and Computer Engineering,
		The University of New Mexico, Albuquerque, NM 87131. 
		E-mail: (venkatesh369@unm.edu; pattichi@unm.edu)}
	\thanks{C. N. Arge is the Chief of the Solar Physics Laboratory in the Heliophysics Science 
		Division at the National Aeronautics and Space Administration\textquotesingle s Goddard Space Flight 
		Center. E-mail: (charles.n.arge@nasa.gov).}
}

\markboth{IEEE Transactions on Image Processing}%
{Image Processing Methods for Coronal Hole Segmentation, Matching, 
        and Map Classification}
%



\maketitle
\begin{abstract}
The paper presents the results from a multi-year effort
   to develop and validate image processing methods for
   selecting the best physical models based on
   solar image observations.
The approach consists of
    selecting the physical models based
    on their agreement with coronal holes extracted
    from the images.
    
Ultimately, the goal is to use
    physical models to predict
    geomagnetic storms.

We decompose the problem into three
  subproblems:
  (i) coronal hole segmentation based on physical constraints,
  (ii) matching clusters of coronal holes between different maps, and
  (iii) physical map classification.

For segmenting coronal holes,
  we develop a multi-modal method that 
  uses segmentation maps from 
  three different methods to initialize
  a level-set method that
  evolves the initial coronal hole segmentation
  to the magnetic boundary.  
Then, we introduce a new method
  based on Linear Programming for
  matching clusters of coronal holes.
The final matching is then performed using Random Forests.
  
The methods were carefully validated using
  consensus maps derived from multiple readers,
  manual clustering, manual map classification, and   
method validation for 50 maps.
The proposed multi-modal 
   segmentation method significantly outperformed
   SegNet, U-net, Henney-Harvey, and FCN
   by providing accurate boundary detection.
Overall, the method gave a 95.5\% 
  map
  classification accuracy.
\end{abstract}


\section{Introduction}
Intense solar activity can cause severe disruptions
   to the Earth's magnetic field.
Typically, open magnetic field lines on the sun 
   form the origin of solar wind that reaches 
   the earth.   
Forecasting requires accurate physical modeling
   of the sun's magnetic field and accurate
   tracking of open magnetic field lines.   
Solar regions characterized by open magnetic
   lines are associated with coronal holes.
Thus, in summary, to support forecasting,
   our goal is to develop
   solar image analysis methods that 
   can automatically segment coronal holes in 
   solar observations, match them to coronal holes generated
   by candidate physical models, and then
   select the best candidate models for forecasting.

\begin{figure}[!tb]
 \includegraphics[width=\linewidth]{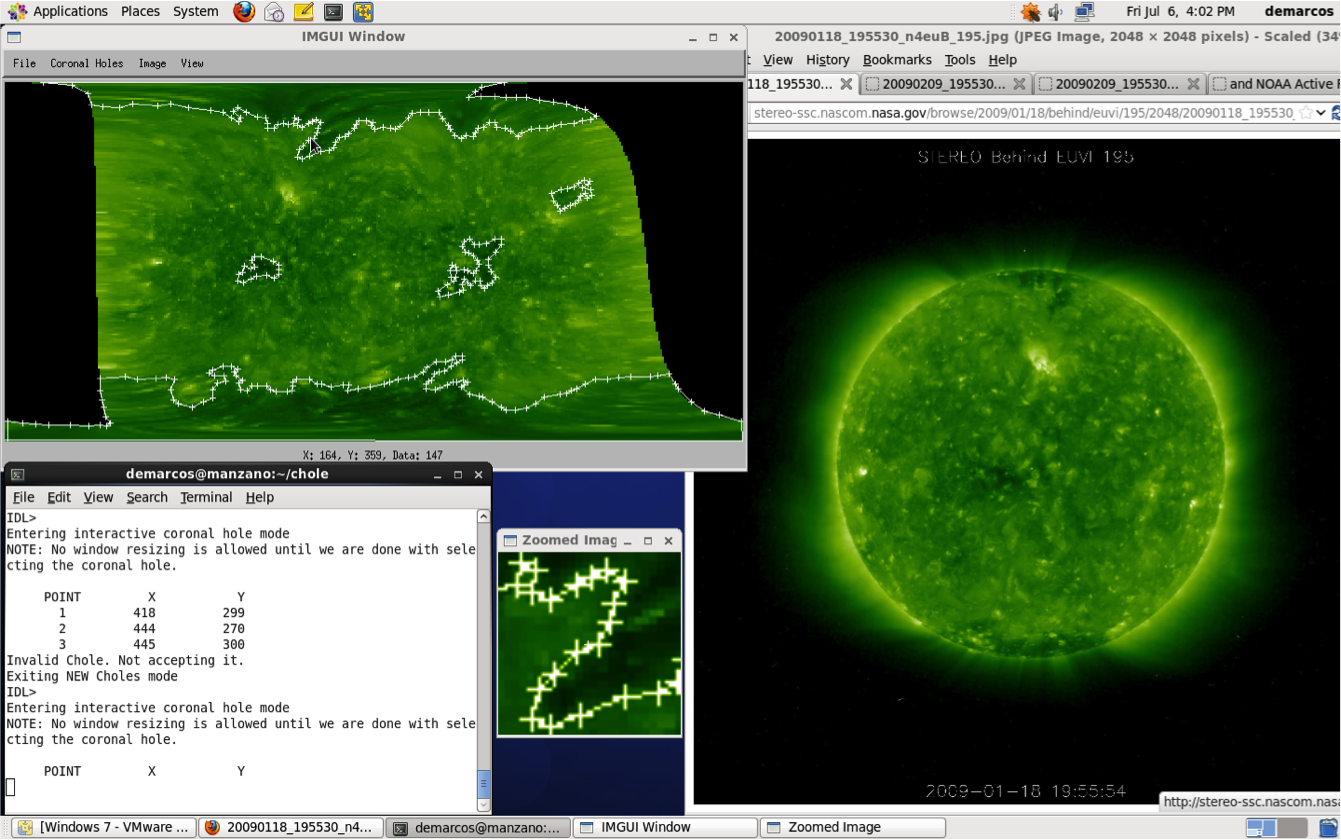}
 \caption{The coronal hole segmentation problem.
 	The figure illustrates the manual segmentation process where
 	    the coronal holes appear as dark regions in
 	    the Extreme ultraviolet (EUV) images.  
 	Black regions represent solar regions for which
 	    we do not have any observations.    	
     }
	\label{fig:idl_tool}
\end{figure}

We present an example of the coronal hole
   segmentation problem in Fig. \ref{fig:idl_tool}.
Coronal holes are distinguished as regions of
   the same magnetic polarity, lower electron density, 
   and lower temperatures \cite{munro1972properties}.
In the extreme ultraviolet (EUV) region,
   coronal holes appear darker than their surrounding regions
   (see Fig. \ref{fig:idl_tool}).

Physical models of the corona and solar wind include Potential Field Source
        Surface based models, such as the Wang-Sheeley-Arge (WSA) model\cite{arge2004stream}
        and the magnetohydrodynamics (MHD) models such as CORHEL
        \cite{linker1999magnetohydrodynamic}.
Physical models use
   solar image observations of the photospheric magnetic
   field 
   as their input to model the solar wind propagating throughout the heliosphere. 
Unfortunately, our physical models are incomplete.
Certain parameters are missing. 
Examples of missing parameters include  
  (a) the source surface height in the Potential Field Source Surface (PFSS) model 
  (see references in \cite{arge2004stream}) 
  of the solar atmosphere (corona), and (b) the amount of heating in more advanced Magnetohydrodynamics (MHD) coronal models. 
In what follows, we assume the use of the WSA model based on PFSS.
In WSA, the amount of open flux and hence the coronal hole area provided by 
  the physical model depends on the source surface radius used (i.e., it is a free parameter in the model beyond which all magnetic field lines flow out into space forming the solar wind). 
More advanced MHD models estimate the amount of heating that will in-turn 
  affect the amount of open flux and hence the coronal hole area 
  \cite{Arge2000, Arge2003, Arge2004}.    
  
To assess the performance of a map produced by a model, there is a need
to establish ground truth from solar observations. 
Unfortunately, the standard practice based on segmentations by a single
     human expert can produce significant inter-rater variability.
To reduce inter-rate variability, we require that the solar images
      should first be manually segmented by two independent experts.
After manual segmentation, the experts meet and agree upon
      a consensus map that represents their combined efforts.

\begin{figure}[!t]
	\centering
	\subfloat[Polar coronal hole matching with strong overlap.]
	{
		\includegraphics[width=0.45\linewidth]{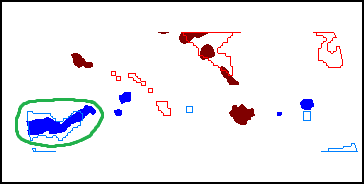}
		\label{subfig:PtoP_Similar}
	}
	~\subfloat[Polar coronal hole matching with wrap-around effect.]
	{
		\includegraphics[width=0.45\linewidth]{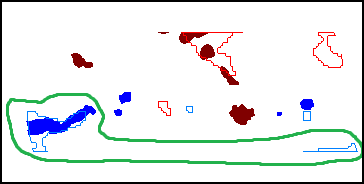}
		\label{subfig:PtoP_Dissimilar}
	}	

	\subfloat[Mid-latitude matchings with good overlap.]
	{
		\includegraphics[width=0.45\linewidth]{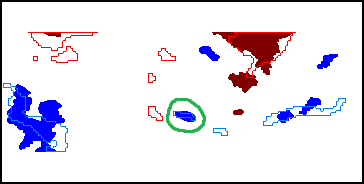}
		\label{subfig:MtoM_Similar}
	}
	~\subfloat[Mid-latitude matching with weak area overlap and
		strong position matching.]
	{
		\includegraphics[width=0.45\linewidth]{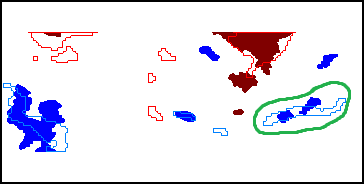}
		\label{subfig:MtoM_Dissimilar}
	}
	
	\subfloat[Polar to mid-latitude matching.]
	{
		\includegraphics[width=0.45\linewidth]{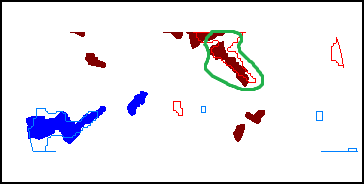}
		\label{subfig:PtoM}
	}
	~\subfloat[Model generated (new) coronal holes that are not seen
	    in the observations.]
	{
		\includegraphics[width=0.45\linewidth]{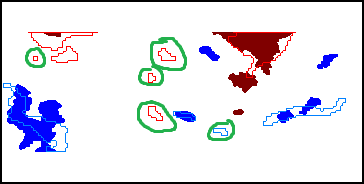}
		\label{subfig:Gen}
	}
	
	\subfloat[Example showing several coronal holes missing from the observations.]
	{
		\includegraphics[width=0.45\linewidth]{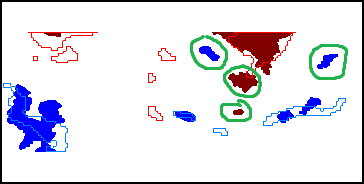}
		\label{subfig:Rem}
	}
	\caption{\label{fig:matching} 
Coronal hole cluster matching problem. The thick green lines are used for circling out the different matching scenarios. The consensus maps are represented by solid colored coronal hole maps. Negative/positive-polarity coronal holes are depicted as blue/red. The hollow, light-blue/red curves represent the boundaries of the negative/positive-polarity coronal holes derived from the physical models. }
\end{figure}

To automate the segmentation process,
   we consider an automatic segmentation approach.
We begin with a review of modern segmentation methods
   and follow-up with a brief summary
   of our proposed approach.
We note that modern
   segmentation methods have been dominated by 
   neural-network based approaches.   
We consider recent segmentation methods
   that are based on convolutional networks
   for visual image classification.
As we shall describe later, 
   we incorporated the best performing methods
   into our segmentation system
   to allow for better initialization
   to our level-set segmentation approach.    
   
In \cite{long2015fully,shelhamer2016fully}, 
  the authors developed fully convolutional
  networks (FCN) for semantic segmentation.
The basic idea is to use transfer learning from
  popular, pre-trained visual classifiers
  (e.g., from AlexNet \cite{AlexNet}, VGG net \cite{VGGNet}, and GoogLe Net \cite{GoogleNet})
  and add multi-scale deconvolutional layers to
  predict ground truth segmentation maps from 
  the previously trained convolutional layers.
The deconvolutional layers are initialized
with linear interpolation kernels
and trained on the specific segmentation
images.
In \cite{long2015fully,shelhamer2016fully}, 
the authors were able to produce state of the
art segmentations by combining the final prediction
layer with 
deconvolutional layers derived from
32x, 16x, and 8x  downsampling strides.   

The most significant limitation of the  
   fully connected neural networks comes from the loss of resolution
   accuracy due to the upsampling operation \cite{shelhamer2016fully}.
To understand the limitation, in the updated
   version of their manuscript \cite{shelhamer2016fully},
   the authors suggest a method to deliver
   upper bounds on performance accuracy.
As an approximate upper bound of what can be achieved,
   the authors took the ground truth segmentation
   images, downsample them by different factors
   and then upsampled them to the original
   resolution.
They then compared the upsampled images
against the unprocessed images to establish
   an approximation of what can be achieved.
Thus, as explored by the creators of the   
   method,
   it is quite clear that such networks should not
   be used to detect small regions (such as small
   coronal holes) or to achieve pixel-level
   resolution accuracy.
For our application, it is also important
   to note that the size of the coronal holes
  may appear to be significantly reduced
  due to missing observations, where the spacecraft
  was unable to collect any data.  

Instead of the FCN approach of combining 
    images from different resolutions,
    the SegNet architecture considered        
    the development of a simpler
    auto-encoding architecture \cite{SegNet}.
Thus, while both FCN and SegNet performed
    very well, a fundamental advantage of the SegNet
    architecture is that it is based on learning
    far fewer parameters.
For SegNet, the encoder is still
    made of convolutional layers followed
    by max pool layers.
However, instead of combining the reduced resolution maps,
    the decoder is made of just the transposed operations.
Thus, the decoder consists of unpool and deconvolution layers
    that are in the reversed-order (compared to the encoder order).
More specifically, the SegNet architecture
    consists of 4 layers with 64 
    feature maps with 7x7 convolution kernels,
    ReLU activation functions,
    followed by max pool layers over 2x2
    non-overlapping regions.
For the decoder, the max-sample locations are used
to reverse each layer, starting with the last layer
and continuing to the first layers, without 
the use of any activation functions.
The final layer consists of SoftMax classification
   into different categories.
A Bayesian extension of SegNet appeared in 
   \cite{BayesianSegNet}.
In \cite{BayesianSegNet}, the authors made it very
   clear that there is strong model uncertainty at
   the boundaries.
Hence, as we shall demonstrate, post-processing
   SegNet results by a carefully designed
   level-set method can yield accurate 
   boundary segmentation.

For the the U-net architecture, the authors
   followed a hybrid approach \cite{UNet}.
As for SegNet, the decoder includes
   the transpose operations.
However, similar to FCN, the decoder combines
   the decoded outputs with the corresponding
   encoder outputs from each one of the convolution layers
   at both the original and reduced-resolution layers.
For U-net, the number of parameters to be learned
   is much larger than for SegNet, but it is also significantly
   less than FCN.
As for SegNet and FCN, U-net cannot perform accurate boundary segmentation.   

As mentioned earlier,
   we propose a new approach
   that uses the best-performing algorithms to initialize
   a level-set segmentation method.
The basic idea is to initialize
   level-set segmentation close
   to the coronal-hole boundaries,
   and then allow level-sets to evolve
   the coronal-hole boundary curves to 
   the magnetic boundaries.
To accomplish this, the level-set function
   is modified to take into account
   both the magnetic boundary as well as
   the local gradient of the EUV image.   
Furthermore, to reduce artifacts,
   we post-process the initial segmentation
   maps to remove invalid coronal holes
   that may not be
   unipolar or appear dark.

Based on the segmented images, our goal is to determine 
      the physical map models that best reflect the observations.
Then, as mentioned earlier,
      space weather forecasting is based on the use of physical maps
      to predict disruptions to the earth's magnetic field.
To accomplish this task, we introduce the coronal hole matching problem
     as an intermediate step.

We demonstrate the coronal hole matching problem in Fig. \ref{fig:matching}.
Each example in Fig. \ref{fig:matching} represents
     a \textit{consensus map} (solid red and blue regions)
     and the physical model map (hollow and light red and blue regions).              
Coronal hole matching poses several challenges.
First, we have wrap-around effects since the coronal holes
    are located on a spherical surface
    (see Fig. \ref{fig:matching}(a) and \ref{fig:matching}(b)).
Second, coronal hole areas and distances also require geometric corrections.
Third, matching needs to consider missing pixel observations
    (see black pixels in Fig. \ref{fig:idl_tool}),
    the generation of new coronal holes by the physical models
    (see Fig. \ref{fig:matching}(f)),
    or the fact that some of the coronal holes may be missing entirely from
    the physical models (see Fig. \ref{fig:matching}(g)).
Fourth, due to randomization of parameters in the physical model
    and the fact that observations may come from different times,
    there is strong variability in the number, location, and
    the areas of the coronal holes.    
Thus, instead of matching individual coronal holes,
    we decided to cluster them together and
    match their clusters (see Fig. \ref{fig:matching}).        

We summarize the primary contributions of the paper into:
\begin{itemize}
  \item Development of a new dataset segmentation problem based
              on consensus maps derived from two independent experts.              
           %
 \item A new segmentation method for detecting coronal holes based on
 	a level-set method initialized 
 	by a combination of the Henney-Harvey algorithm \cite{Henney2005}, SegNet \cite{SegNet},
 	and FCN \cite{long2015fully,shelhamer2016fully}.
 	The proposed approach achieves substantial improvements over
 	    prior methods.
 \item Development of a new approach for matching,
           and detecting missing or the appearance of new coronal holes.
          Due to the significant physical constraints, our approach
            is significantly different from any prior method.
          For example, in  \cite{Chen2006, Li2010, Dewan2011}, the authors
           studied a similar problem in cells and nuclei matching.
         These approaches used heuristics based on shape, size, and
            empirical set distances.
         In our case, we develop a Linear Programming model using
            spherical geometry to compute
            physical distances and areas derived from their projections.
       Furthermore, unlike cells, we have the new problems of having
            to detect the random creation and disappearance of coronal holes.
   \item Development of a physical map classification system.
       This new system is built upon the coronal hole segmentation and matching
            approaches to determine the physical maps that will be used for forecasting.         
\end{itemize}

The rest of the paper is organized into four sections. Section \ref{sec:sec2_manualClassification}
describes carefully developed manual protocols that help in creating consistent ground truth for
segmentation, matching and classification. Following this we have Section \ref{sec:methodology},
giving a brief introduction to segmentation and a detailed description of matching and classification
algorithms. In Section \ref{sec:results} we go over results followed by conclusions in Section \ref{sec:conclusion}.
\section{Solar image analysis problem setup
             \label{sec:sec2_manualClassification}}
\subsection{Coronal hole segmentation problem setup}\label{sec:hole}

To generate the consensus map, we followed a three-step process.
First, a human reader was trained to manually outline
the individual coronal holes based on
physical constraints 
(e.g., unipolarity and appearance, see Fig. \ref{fig:idl_tool}).
Second, the process was repeated with a second human reader.
Third, the two human readers got together to generate the consensus maps
by agreeing on what constitutes a coronal hole based on a second look
and by reviewing their original maps.
\subsection{Coronal hole matching problem setup}\label{sec:match}
In this section, we provide a summary of the 
   process of creating ground truth for the 
   coronal hole matching algorithm.
We summarize the manual protocol for clustering
   coronal holes in Fig. \ref{fig:holeClusterManual}
   and demonstrate the application of the protocol
   in Fig. \ref{fig:clusteringInAction}.
   
\begin{figure}[!t]
        \begin{itemize}
                \item[{\bf CR1.}] \textbf{Cluster polar coronal holes}:
                   Coronal holes that are within $30^{\circ}$ from north and south poles are clustered together.
                   Furthermore, any coronal hole that crosses the $30^{\circ}$ lines is clustered into
                       the corresponding polar coronal hole.
                       The resulting clusters represent the north and south polar coronal hole clusters.
                \item[{\bf CR2.}] \textbf{Nearby clustering}: Coronal holes that are extremely close to each other
                are clustered together.
                \item[{\bf CR3.}] \textbf{Small-small clustering}: 
                Groups of small coronal holes that are relatively close to each other are
                clustered together.
                \item[{\bf CR4.}] \textbf{Large-small clustering}:
                A small coronal hole that is close to a much larger one is considered
                part of the larger cluster that involves the larger coronal hole.
                \item[{\bf CR5.}] \textbf{No large-large clustering}:
                In general, larger coronal holes are not clustered together
                unless they are extremely close to each other (see \textbf{CR2}).
        \end{itemize}
        \caption{Coronal hole clustering protocol. The manual protocol was
            used for implementing physically meaningful and consistent rules.
        The protocol was applied several times to
        ensure reproducibility and little to no intra-rater variability.}
    \label{fig:holeClusterManual}
\end{figure}
\begin{figure}[!b]
	\subfloat[\textbf{CR1. Cluster polar coronal holes.}]
	{
		\includegraphics[width=0.45\linewidth]{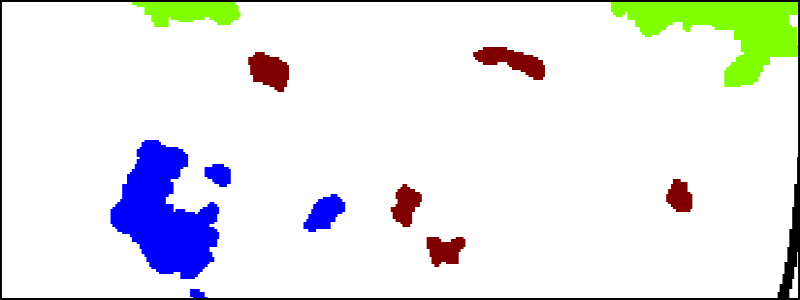}
		\label{subfig:cr1a}
	}
	\subfloat[\textbf{CR1. Cluster polar coronal holes.}]
	{
		\includegraphics[width=0.45\linewidth]{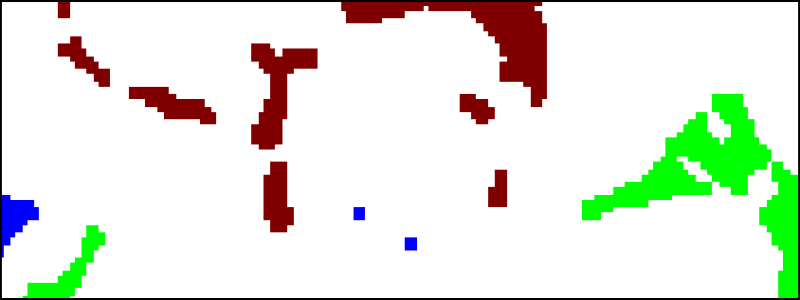}
		\label{subfig:cr1b}
	}
	
	\subfloat[\textbf{CR2. Nearby clustering}.]
	{
		\includegraphics[width=0.45\linewidth]{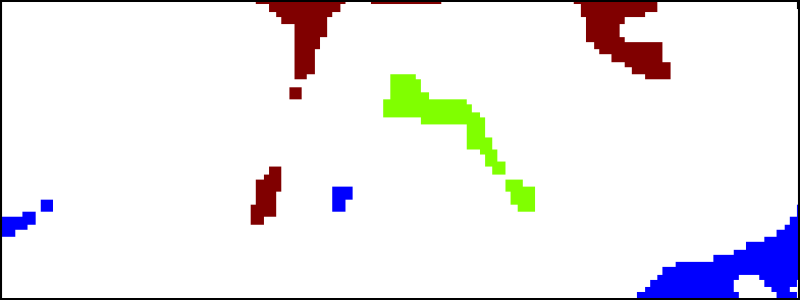}
		\label{subfig:cr2}
	}
	\subfloat[\textbf{CR3. Small-small clustering.}]
	{
		\includegraphics[width=0.45\linewidth]{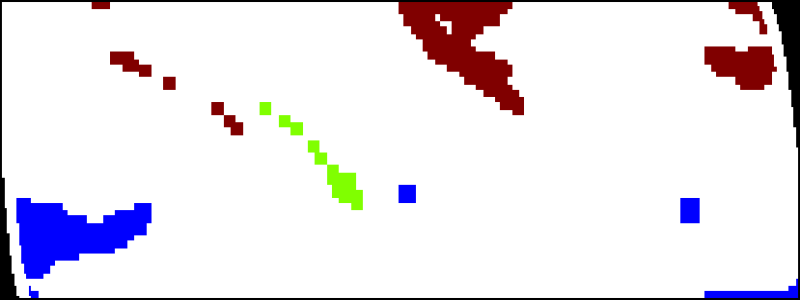}
		\label{subfig:cr3}
	}
	
	\subfloat[\textbf{CR4. Large-small clustering.}]
	{
		\includegraphics[width=0.45\linewidth]{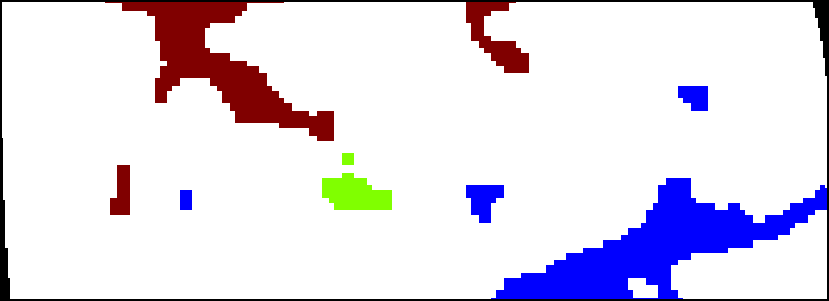}
		\label{subfig:cr4}
	}
	\subfloat[\textbf{CR5. No large-large clustering.}]
	{
		\includegraphics[width=0.45\linewidth]{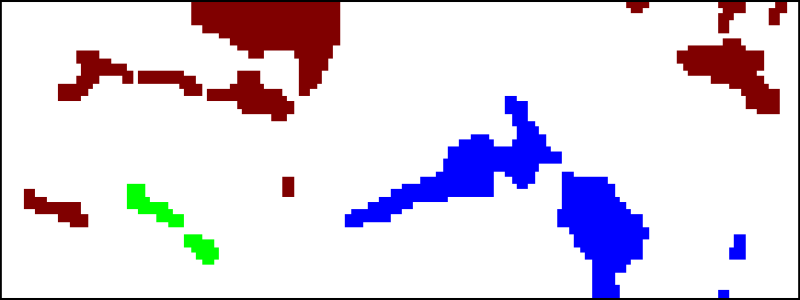}
		\label{subfig:cr5}
	}
	\caption{Manual application of coronal holes clustering rules that
		meet physical requirements. The green coronal holes are combined
		into a single cluster. Here, CR1 to CR5 refer to the rules
		that are summarized in Fig. \ref{fig:holeClusterManual}.
	 We use red to depict coronal holes with positive polarity.
	 We use blue to depict coronal holes with negative polarity.	
	 }
	\label{fig:clusteringInAction}
\end{figure}

We describe a manual protocol that ensures reproducibility
    of the coronal hole matching problem in
    Fig.  \ref{fig:clusterMatchingManual}.
The protocol is used to match clusters of the same polarity.
The remaining clusters that could not be matched,
     depending on their presence in the consensus maps,
     are labeld as new or missing
     (e.g., see  Fig. \ref{fig:matching} ).
     
\begin{figure}[!t]
\begin{itemize}
        \item[{\bf M1.}] {\bf Polar to polar matching:} Polar clusters with a relatively large area
        overlap (70\% to 100\%) are matched.
        \item[{\bf M2.}] {\bf Polar to mid-latitude matching:}
        A coronal hole cluster from the consensus map that is located
        in the mid-latitude region is matched to a polar cluster from
        the physical model when they overlap by at-least 15\% to 20\%, or more.
        \item[{\bf M3.}] {\bf Mid-latitude to mid-latitude matching:}
        Mid-latitude clusters are matched with good area overlap 
        (e.g., overlap area $>30\%$) or weaker area overlap but
        good localization.
\end{itemize}
        \caption{Cluster matching protocol. The manual protocol was used
        to produce a reference matching approach for training and
    testing matching rules. The protocol was applied several times to
    ensure reproducibility and little to no intra-rater variability.}
\label{fig:clusterMatchingManual}
\end{figure}

\begin{figure}[!b]
        \begin{algorithmic}[1]    
                \State \textit{\textbf{Group}} maps based on matched clusters
                \State \textit{\textbf{Rank}} groups as rank 1 (better) or rank 2 (worse).      
                \Statex 
                \If {(both ranks represent good matches)}
                \State \textit{\textbf{Classify}} all maps of the day as good matches.
                \EndIf
                \Statex
                \If {(both ranks represent bad matches)}
                \State \textit{\textbf{Classify}} all maps of the day as bad matches.
                \EndIf 
                \Statex
                \If {(rank 1 is acceptable and not rank 2)}
                \State \textit{\textbf{Classify}} maps of rank 1 as good matches.
                \State \textit{\textbf{Classify}} maps of rank 2 as bad matches.
                \EndIf      
        \end{algorithmic}
        \caption{
        Map classification protocol to ensure consistency.
        For consistent classification, physical maps are grouped together
        based on their overall matching performance.
        Each group is then ranked.
        Classification is based on the group rank (good or bad).}
        \label{fig:classAlg}
\end{figure}

\begin{figure}
	\subfloat[Model 1 is in group A (rank=2, bad match).]
	{
		\includegraphics[width=0.45\linewidth]{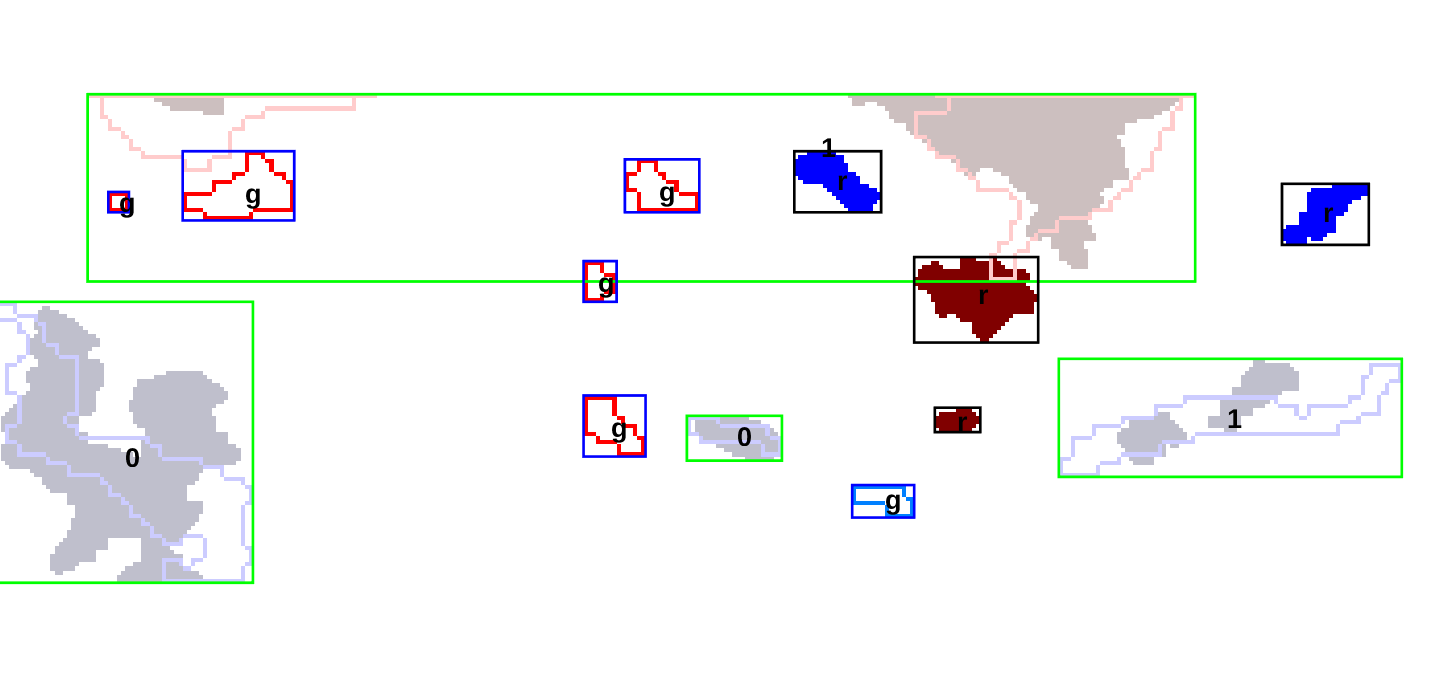}
		\label{subfig:m1g1}
	}
	~\subfloat[Model 6 is in group A (rank=2, bad match).]
	{
		\includegraphics[width=0.45\linewidth]{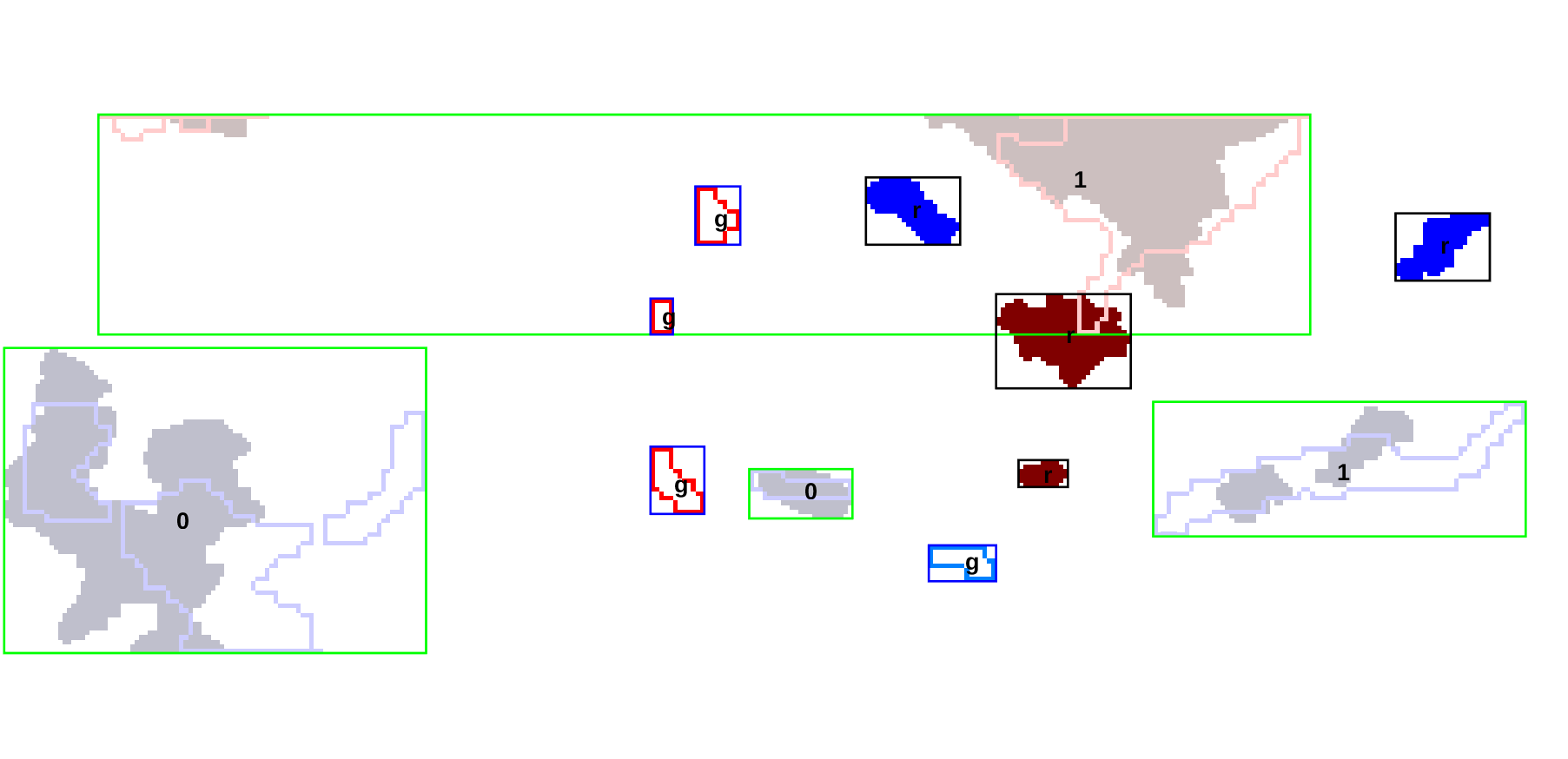}
		\label{subfig:m6g1}
		
	}
	
	\subfloat[Model 8 is in group A (rank=2, bad match).]
	{
		\includegraphics[width=0.45\linewidth]{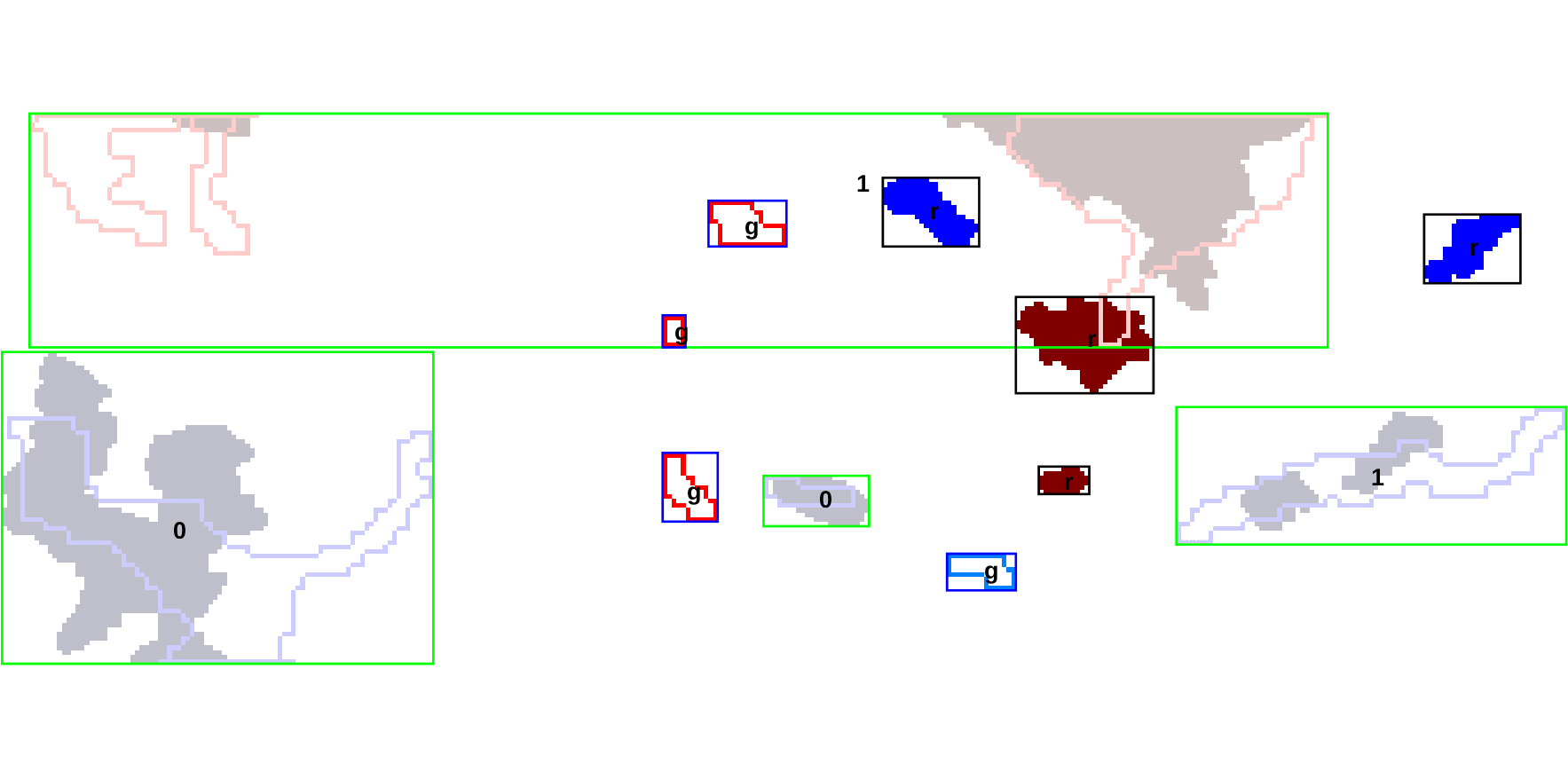}
		\label{subfig:m8g1}
		
	}
	~\subfloat[Model 9 is in group A (rank=2, bad match).]
	{
		\includegraphics[width=0.45\linewidth]{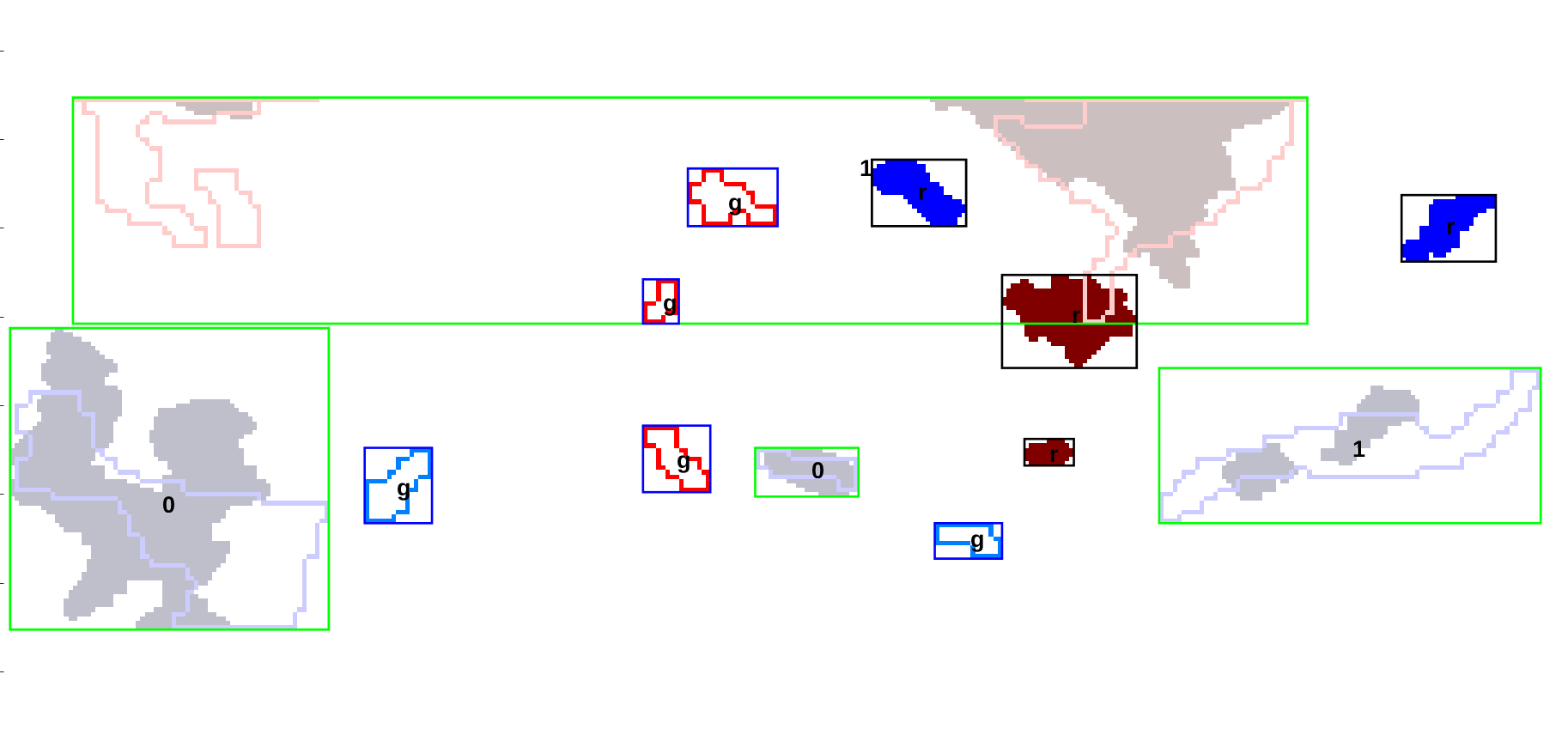}
		\label{subfig:m9g1}
		
	}
	
	\subfloat[Model 11 is in group A (rank=2, bad match).]
	{
		\includegraphics[width=0.45\linewidth]{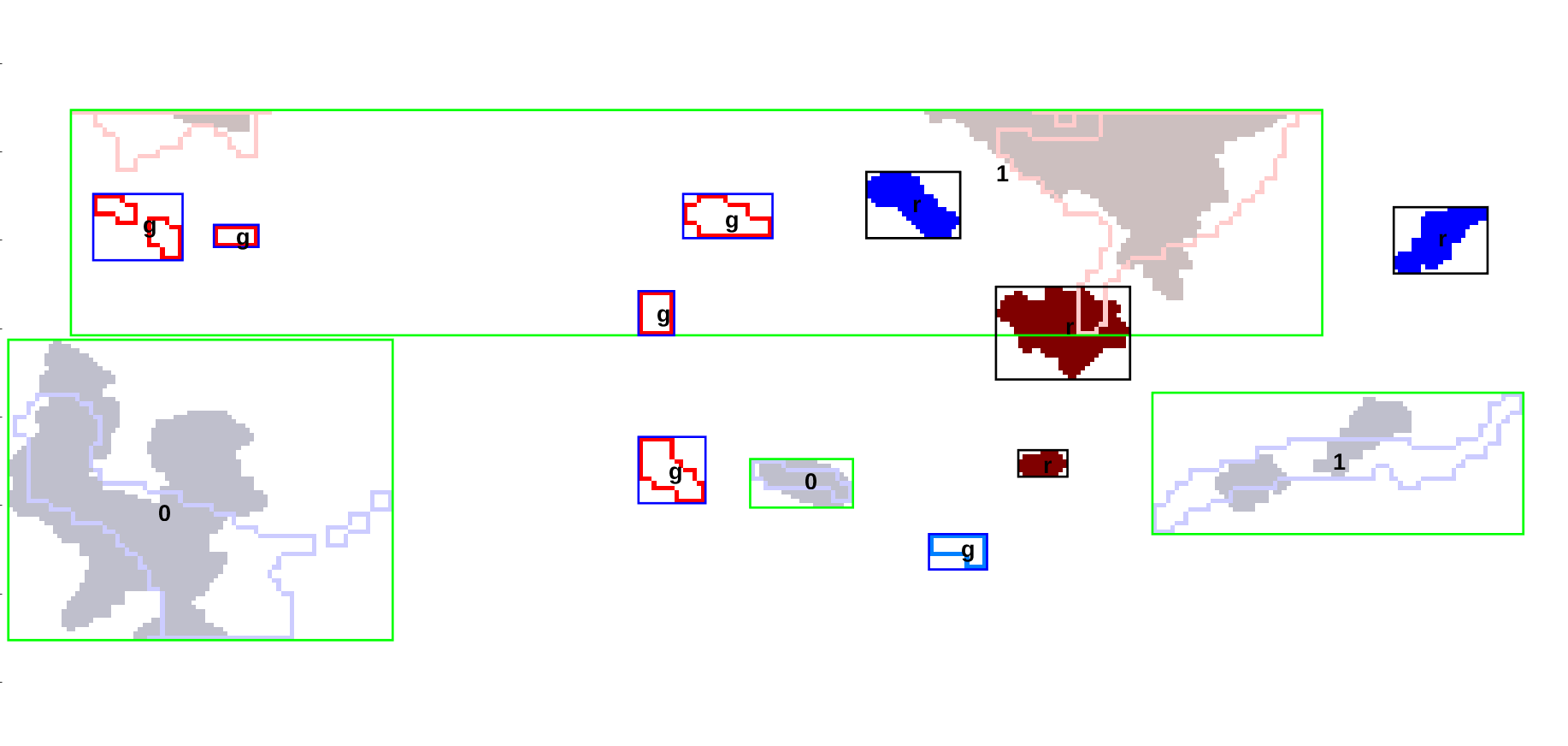}
		\label{subfig:m11g1}
		
	}
	~\subfloat[Model 4 is in group B (rank=1, good match).]
	{
		\includegraphics[width=0.45\linewidth]{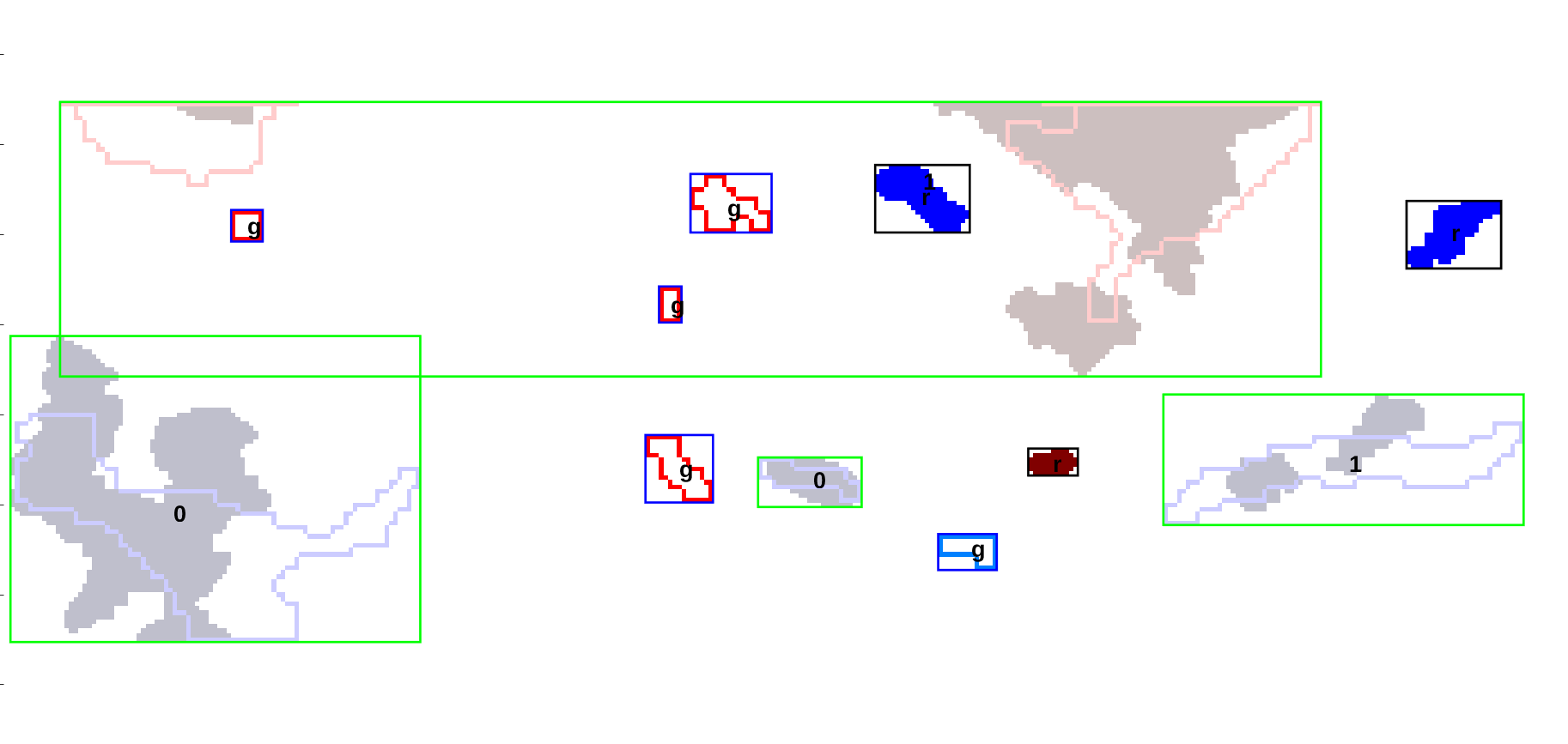}
		\label{subfig:m4g2}
		
	}
	
	\subfloat[Model 10 is in group B (rank=1, good match).]
	{
		\includegraphics[width=0.45\linewidth]{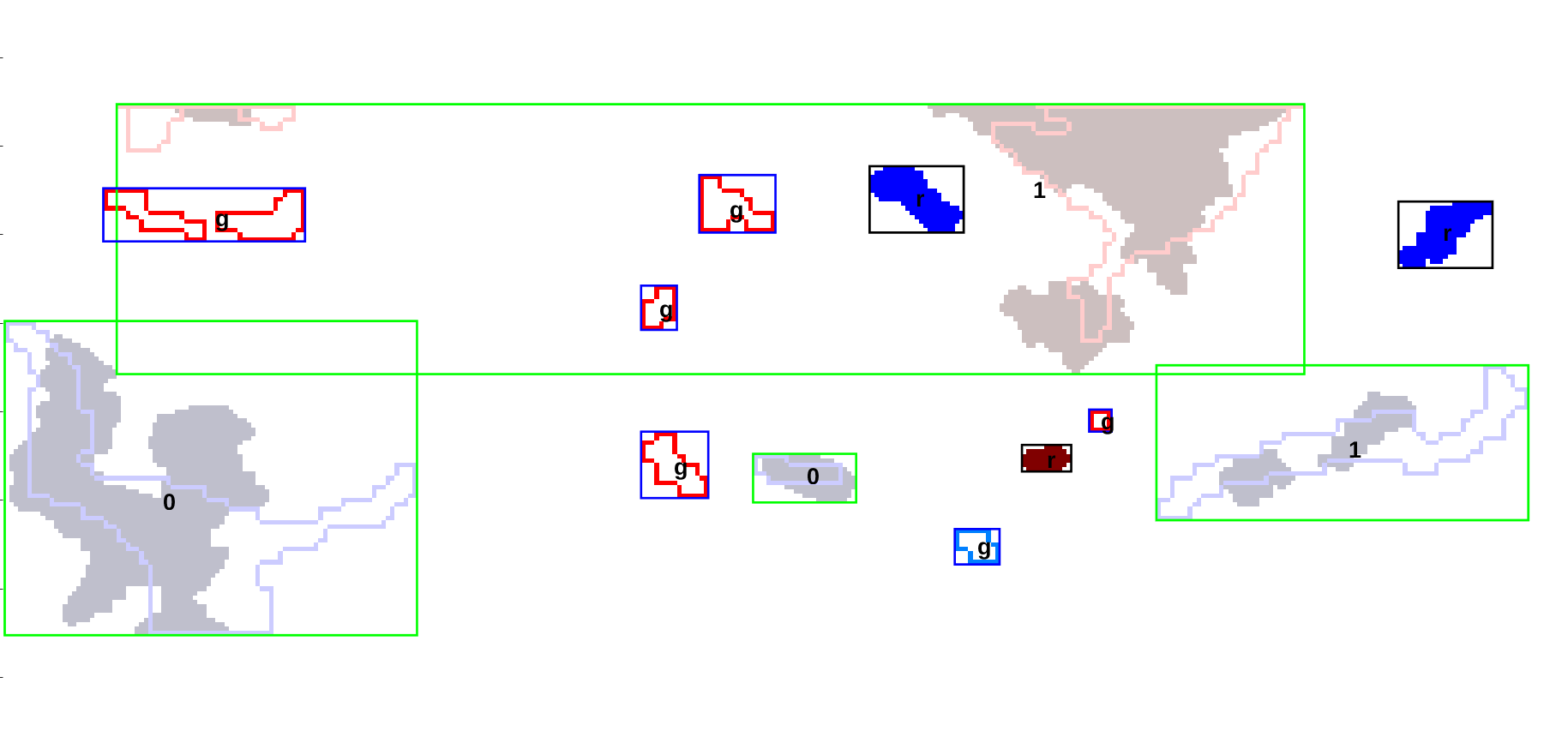}
		\label{subfig:m10g2}
	}
	~\subfloat[Model 12 is in group B (rank=1, good match).]
	{
		\includegraphics[width=0.45\linewidth]{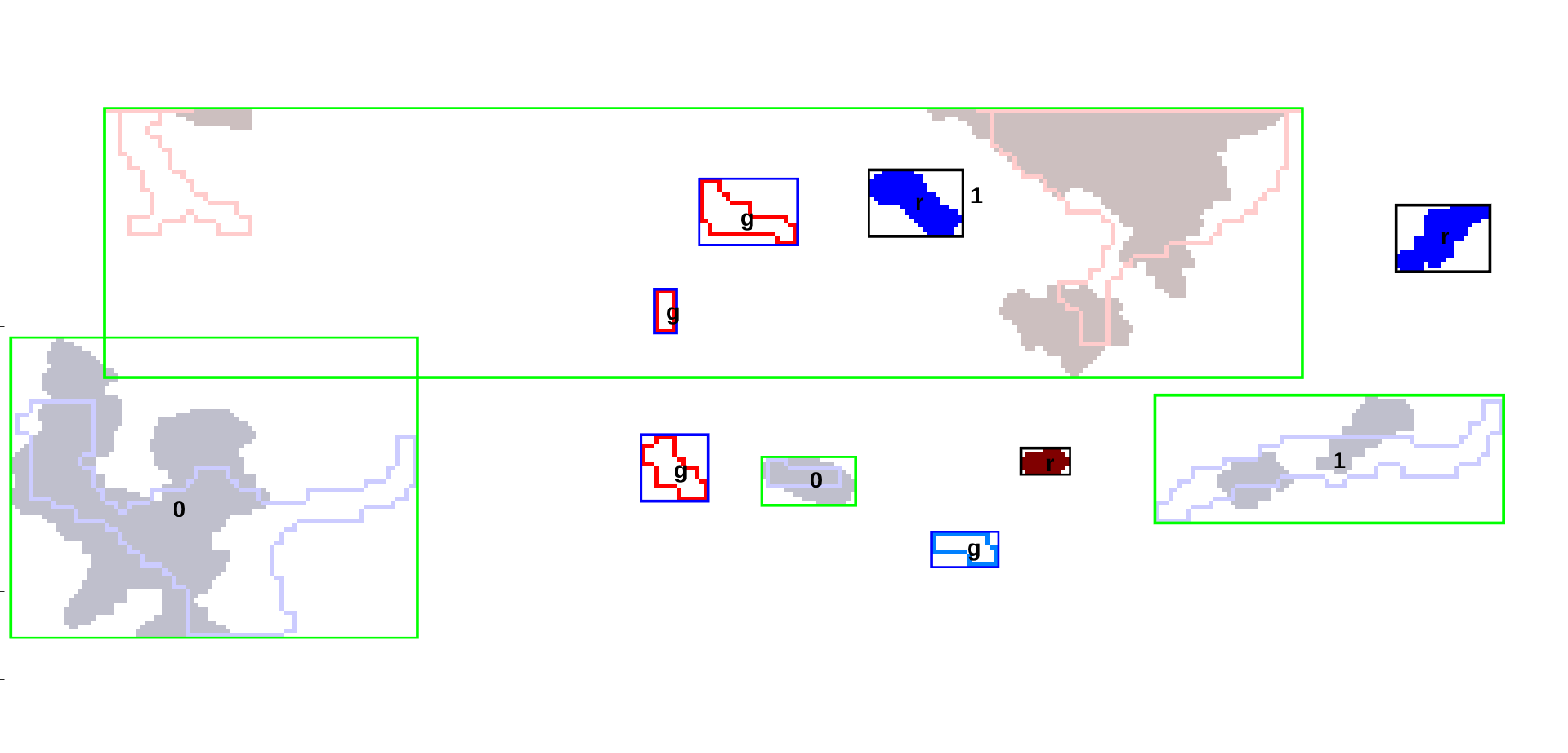}
		\label{subfig:m12g2}
		
	}
	\caption{
	Map classification example (see Fig. \ref{fig:classAlg}).
	The maps are grouped into groups A and B.
	Group A is classified as rank 2.
	Group B is classified as rank 1. 
	Maps with rank 1 are classified as good.
	Maps with rank 2 are classified as bad. 
    }
	\label{fig:classificationInAction}
\end{figure}

\subsection{Coronal hole map classification problem setup}\label{sec:map}
To ensure reproducibility of the process, we need to take into account that
    many physical models appear very similar to each other.
Thus, to standardize our approach, we group together
    maps that are virtually indistinguishable and 
    classify each group as opposed to classifying individual maps.

Maps are pre-classified into two
groups. We use ranks to describe each group. 
In the rank 1 group, we include maps that tend to be closer to the consensus map.
In the rank 2 group, we include maps that tend to be further away from the consensus map.
We then make the final classifications of what
constitutes a good and a bad map based on the rules given in Fig. \ref{fig:classAlg}.

To decide the rankings, we examine the mid-latitude coronal holes.
Initially, similar to clustering, we group maps based on how similar
they are to each other.
The collection of all of the groups are then classified 
as being closer to the consensus map (rank 1)
or further from the consensus map (rank 2).
Here, we classify a group as being closer to the consensus map if
it contains a substantial number of matched, 
fewer cases of new (generated) and missing (removed) coronal holes.
A ranked group of maps (rank 1 or 2) is then classified a \textit{good match}
if it is in good agreement with the consensus map, where we also
allow slight over-estimation of the area of the coronal holes.
A group of maps that is not considered a \textit{good match} is
classified as a \textit{bad match}. 
We present a classification example in Fig. \ref{fig:classificationInAction}.

We show examples from two groups in Fig. \ref{fig:classificationInAction}.
Group A maps do not have a matching for the positive polarity coronal hole located 
in the upper-right region (depicted as bright red).
Group B maps do have a matching cluster for the same coronal hole
(depicted as faded red).
Furthermore, group 2 maps missed (removed) fewer coronal holes
(depicted as solid blue here).
Thus, group B maps are thus classified as rank 1 and group A maps 
are classified as rank 2.
Furthermore, since rank 1 maps detected more of the coronal holes,
  rank 1 maps were classified as \textit{good matchings}.
On the other hand, since rank 2 maps missed some of the coronal holes,  
  rank 2 maps were classified as \textit{bad matchings}.

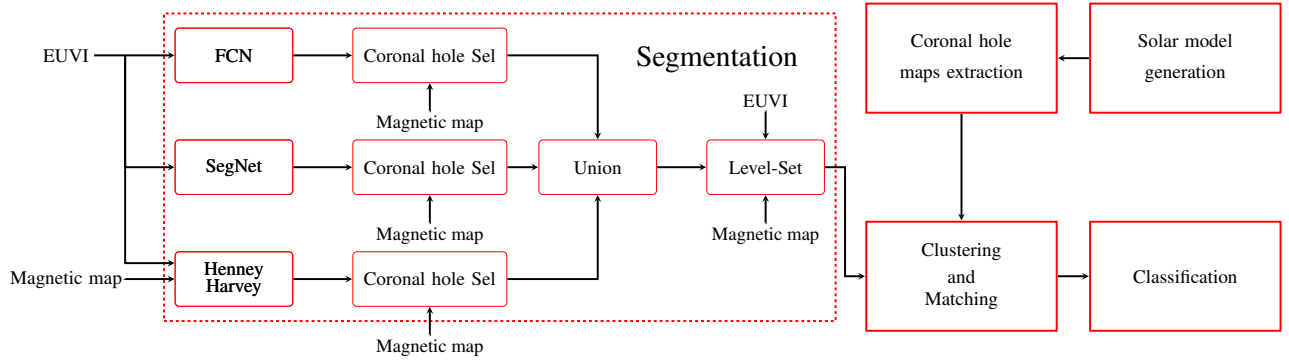
\begin{figure*}[!t]  
	\centering
	\resizebox{0.95\textwidth}{!}{%
		\begin{tikzpicture}[node distance=2cm]
		
		\node(syn)[EMPTY, draw=none, text width=2.5cm, text centered, 
		minimum width = 1.5cm, minimum height = 1.5cm, scale=1.3, text width = 3cm]
		{
			EUVI
		};
		\node(mag)[EMPTY, text width=2.5cm, text centered,below of =syn, node distance = 8cm,
		rounded corners, minimum width = 1.5cm, minimum height = 1.5cm, scale=1.3, text width=4cm]
		{
			Magnetic map
		};

		\node(fcn)[BOXL3, rounded corners, node distance = 6cm, right of= syn,minimum width = 1.5cm, minimum height = 1.5cm, scale=1.3]
		{
			FCN
		};
		
		\node(sn)[BOXL3, below of = fcn, rounded corners, node distance = 4cm, minimum width = 1.5cm, minimum height = 1.5cm, scale=1.3]
		{
			SegNet
		};
		\node(hh)[BOXL3, below of = sn, rounded corners, node distance = 4cm, minimum width = 1.5cm, minimum height = 1.5cm, scale=1.3]
		{
			Henney\\
			Harvey
		};

		\node(fcn)[BOXL3, rounded corners, node distance = 6cm, right of= syn,minimum width = 1.5cm, minimum height = 1.5cm, scale=1.3]
		{
			FCN
		};
		
		\node(sn)[BOXL3, below of = fcn, rounded corners, node distance = 4cm, minimum width = 1.5cm, minimum height = 1.5cm, scale=1.3]
		{
			SegNet
		};
		\node(hh)[BOXL3, below of = sn, rounded corners, node distance = 4cm, minimum width = 1.5cm, minimum height = 1.5cm, scale=1.3]
		{
			Henney\\
			Harvey
		};

		\node(fcnsel)[BOXL3, right of = fcn, rounded corners, node distance = 7cm, minimum width = 4cm, text width=4cm, minimum height = 1.5cm, scale=1.3]
		{
			Coronal hole Sel
		};
		\node(snsel)[BOXL3, right of = sn, rounded corners, node distance = 7cm, minimum width = 4cm, minimum height = 1.5cm, scale=1.3, text width=4cm,]
		{
			Coronal hole Sel
		};
		\node(hhsel)[BOXL3, right of = hh, rounded corners, node distance = 7cm, minimum width = 4cm, minimum height = 1.5cm, scale=1.3, text width=4cm]
		{
			Coronal hole Sel
		};

		\node(union)[BOXL3, right of = snsel, rounded corners, node distance = 6cm, minimum width = 1.5cm, minimum height = 1.5cm, scale=1.3]
		{
			Union
		};

		\node(ls)[BOXL3, right of = union, rounded corners, node distance = 6cm, minimum width = 1.5cm, minimum height = 1.5cm, scale=1.3]
		{
			Level-Set
		};

		\node [ BOXLD2, minimum width = 24cm, minimum height = 11cm, fit={(fcn) (sn) (hh) (fcnsel) (snsel) (hhsel) (ls)}](segbox) {
			\vspace{7cm}
			\hspace{15cm}
			Segmentation
		};

		\node(cm)[BOXL2, right of = ls, node distance = 7cm,
		minimum width = 5cm, minimum height = 3cm, scale=1.3,yshift=-30mm, text width=5cm]
		{
			Clustering \\[0.25cm]and \\[0.25cm]Matching 
		};
		\node(cls)[BOXL2, right of = cm, node distance = 8cm,
		minimum width = 5cm, minimum height = 3cm, scale=1.3, text width=5cm]
		{
			Classification
		};

		\node(me)[BOXL2, right of = ls, node distance = 7cm,
		minimum width = 5cm, minimum height = 3cm, scale=1.3,yshift=30mm, text width=5cm]
		{
			Coronal hole \\[0.5cm] maps extraction
		};
		
		\node(mg)[BOXL2, right of = me, node distance = 8cm,
		minimum width = 5cm, minimum height = 3cm, scale=1.3, text width=5cm]
		{
			Solar model \\[0.5cm]generation	
		};

		\draw[ARROW, line width=2pt] (1,0) |- (fcn.west) ;
		\draw[ARROW, line width=2pt] (syn.east) to [|-] (sn.west);
		\draw[ARROW, line width=2pt] (syn.east) to [|-] (hh.165);
		\draw[ARROW, line width=2pt] (2.1,-8) |- (hh.west);
		\draw[ARROW, line width=2pt] (fcn.east) to [|-] (fcnsel.west);
		\draw[ARROW, line width=2pt] (sn.east) to [|-] (snsel.west);
		\draw[ARROW, line width=2pt] (hh.east) to [|-] (hhsel.west);
		\draw[ARROW, line width=2pt] (13,-2) node[below, font=\fontsize{19}{0}\selectfont]{Magnetic map} -| (fcnsel.south);
		\draw[ARROW, line width=2pt] (13,-6) node[below, font=\fontsize{19}{0}\selectfont]{Magnetic map} -| (snsel.south);
		\draw[ARROW, line width=2pt] (13,-10) node[below, font=\fontsize{19}{0}\selectfont]{Magnetic map} -| (hhsel.south);
		\draw[ARROW, line width=2pt] (fcnsel.east) -| (union.north);
		\draw[ARROW, line width=2pt] (snsel.east) -- (union.west);
		\draw[ARROW, line width=2pt] (hhsel.east) -| (union.south);
		\draw[ARROW, line width=2pt] (union.east) -- (ls.west);
		\draw[ARROW, line width=2pt] (25,-2) node[above, font=\fontsize{19}{0}\selectfont]{EUVI} -| (ls.north);
		\draw[ARROW, line width=2pt] (25,-6) node[below, font=\fontsize{19}{0}\selectfont]{Magnetic map} -| (ls.south);
		\draw[ARROW, line width=2pt] (mg.west) -- (me.east);
		\draw[ARROW, line width=2pt] (ls.east) to [-|-] (cm.west);
		\draw[ARROW, line width=2pt] (cm.east) -- (cls.west);
		\draw[ARROW, line width=2pt] (me.south) -- (cm.north);

		\end{tikzpicture}
	}
	\caption{\label{fig:IntroBlkDiag}
		General system diagram.
		A collection of coronal hole maps are generated for
		different physical model parameters.
		The input observations are used to generate a candidate
		coronal hole map.
		A hierarichical clustering and matching algorithm
		is used for matching clusters of coronal holes.
		The final classifier is based on the matching results.		
	}
\end{figure*}

\begin{table}[b!]
	\centering	
	\caption{
		SegNet architecture.
		The total number of parameters to be learned 
		is 0.52M.
		The input is the EUV image and the output is
		a segmentation into three categories
		(no-observation, coronal hole, other).   	       
		Here, Conv-bn-ReLU refers to the combination
		of a convolution layer, batch normalization, and an ReLU
		activation function.
	}
	\label{tab:segnet_arch}
	\begin{tabular}{p{1cm} p{6.5cm}}
		\toprule
		\textbf{Layer(s)} & \textbf{Description}\\
		\midrule
		Enc1	&   64 feature maps of 4 3x3 conv-bn-ReLU 	\\
		Pool1	&	2x2 Max pooling with stride of [2, 2] 
		(2$\times$ downs.).	\\
		Enc2	&	Same as Enc1	\\
		Pool2	&	Same as Pool1 (4$\times$ downs. of original) \\
		Unpool1	&	2x2 Max unpooling (restore at $2\times$ orig. res.)	\\
		Dec1	&	4 3x3 conv-bn-ReLU Layers	\\
		Unpool2	&	Same as Unpool1	(restore original res.) \\
		Dec2	&	Same as Dec1	\\
		SoftMax	&	Pixel-level classification	\\
		\bottomrule
	\end{tabular}
\end{table}

\begin{table}[t!]
	\centering	
	\caption{
		FCN architecture.
		The total number of parameters to be learned is 134.2M.
		The input is the EUV image and the output is the segmented image.
	}
	\label{tab:fcn_arch}
	\begin{tabular}{p{1cm} p{7cm}}
		\toprule
		\textbf{Layer(s)} & \textbf{Description}\\
		\midrule
		Conv1	& 64 feature maps with two 3x3 conv-ReLU \\
		Pool1	& 2x2 Max pooling with stride of [2, 2] (2$\times$ downs.) \\
		Conv2	& 128 feature maps with two 3x3 conv-ReLU\\
		Pool2	& Same as Pool1 (4$\times$ downs. of original)\\
		Conv3	& 256 feature maps with three 3x3 conv-ReLU\\
		Pool3	& Same as Pool1 (8$\times$ downs. of original) \\
		Conv4	& 512 feature maps with three 3x3 conv-ReLU\\
		Pool4	& Same as Pool1 (16$\times$ downs. of original)\\
		Conv5	& Same as Conv4\\
		Pool5	& Same as Pool1 (32$\times$ downs. of original)\\
		FC6	    & 4096 7x7 convs for 512 features\\
		D6   	& 50\% dropout\\						
		FC7	    & 4096 fully-connected layer \\
		D7	    & Same as D6\\
		Fuse1	& Combine 16$\times$ and 32$\times$ downsampled outputs \\
		Fuse2	& Combine 8$\times$, 16$\times$, 32$\times$  downsampled outputs \\
		SoftMax	& Pixel-level classification\\
		\bottomrule
	\end{tabular}
\end{table}

\section{Methodology}\label{sec:methodology}
We provide an overview of the proposed approach in Fig. \ref{fig:IntroBlkDiag}.
We decompose the proposed system into three components:
(i) coronal hole segmentation,
(ii) clustering and matching, and
(iii) classification.
We describe the image segmentation approach in
subsection \ref{sec:Segmentation}.
For matching, we consider each physical map against
the segmented map.
To reduce variability, we form clusters of coronal holes
and attempt to match clusters between maps.
The basic method is given in subsection \ref{sec:ClusteringAndMatching}.
Physical maps are then classified based on
their matching performance
as described in section \ref{sec:Classification}.\\

\subsection{Multimodal Segmentation of Coronal Holes}\label{sec:Segmentation} 
We develop a multi-modal segmentation approach as shown
   in Fig. \ref{fig:IntroBlkDiag}.
The basic idea is to provide an initial segmentation
   map that is then input to a level-set method
   that is designed to 
   provide an accurate segmentation by evolving the
   initial segmentation to the magnetic boundary.
   
To initialize the segmentation approach, we use
   three different methods working
   with three different sets of pixel resolutions.
At the pixel resolution level, we use the original
   Henney-Harvey method \cite{Henney2005}
   that uses pixel classification 
   to select candidate coronal holes that are both
   unipolar and darker in the EUV images.
At the $2\times$ and $4\times$ downsampled resolution levels,
   SegNet reconstructs pixel-level segmentations
   by first encoding the EUV image 
   at 1/2 and 1/4 of the original image resolution
   and then decoding the encoded maps to provide
   classification at the full resolution.
At the $2\times$ to $32\times$ downsampled resolution levels,
   FCN reconstructs pixel-level segmentations by
   first encoding at 1/2 to 1/32 of the original image resolution,
   and then combining $8\times, 16\times, 32\times$ encodings to
   form the predicted segmentation image at the full resolution.
   
For the neural-net based methods,
   we only input the EUV images since they are used
   for visual classification by the astronomers and avoid
   combining them with the magnetic images.
For the magnetic images, 
    note that the training set is rather limited. 
There is no possibility of pre-training since they are very different
   from visual images used in standard datasets,
   and we defer further processing by 
   the coronal-hole selector and level-set methods.   
Here, we also note that we had to learn $134.2$M parameters for FCN
   (pre-trained on VGG-16 \cite{VGGNet,VGG16})
   and $0.52$M parameters for SegNet.
In any case, we also verified these claims experimentally.
As we describe in the results section, the performance
   of SegNet when input with both the magnetic and EUV images 
   deteriorated considerably while there was no observed 
   improvement for FCN.

We provide more details on the specific SegNet and FCN
   architectures in Tables \ref{tab:segnet_arch} and \ref{tab:fcn_arch}.
All of the parameters were learned on the training set,
   which represents a random sample of 70\% of the images.
None of these images overlap with the remaining test set that
   are used in the Results section.
For both methods, we used 50 epochs with a mini-batch size of 7.
For SegNet, we used a learning rate of 0.1 with a momentum set to 0.9.
It took about an hour to train the SegNet on dual NVIDIA GTX 1080 video cards 
    with 2,560 cores and 8GB each.
On the same system, it took about 20 minutes for the pre-trained FCN to be re-trained   
   using a learning rate of $10^{-3}$ and a momentum parameter set to 0.9.

The coronal holes resulting from the initial segmentations
   are classified (selected) to further impose physical constraints
   on their appearance.
Thus, since coronal holes are supposed to be dark in EUV images
   and unipolar in the magnetic images, we use
   histograms of EUV intensity (255 bins) and magnetic flux (40 bins)
   as input features to the classifier.
Furthermore, we add the area of each coronal hole as an input feature
   since the downsampling and upsampling operations
   generate small, noisy estimates of the actual coronal holes.
For classifying each coronal hole, we use a Random Forest for each segmentation method.

To select the parameters for the Random Forest, we further split
    the training set into 70\% training for each classifier model
    and 30\% testing for selecting the best classifier model
    (\cite{ESLII}).
We used the out-of-bag error to determine the best models
    that avoid over-fitting:
    (i)   50 trees with 50 splits for Henney-Harvey,
    (ii)  30 trees with 50 splits for SegNet, and
    (iii) 30 trees with 20 splits for FCN.

After identifying candidate coronal holes, we use a union
    operation to combine the outputs.
Here, we recognize that the union operation will likely
    overestimate the actual coronal holes.
However, given the potentially catastrophic consequences
    of missing an actual coronal hole, we prefer to err on
    the side of providing a slight
    over-estimation as opposed to missing one of them.
The union map is then input to the level-set method
    that can provide an accurate estimation
    of the magnetic boundary.
\subsubsection{Level-set Segmentation}\label{sec:LS}
We develop a multi-modal segmentation approach by expanding
     the Distance Regularized Level Set Evolution (DRLSE) method \cite{chunming2010}
     to account for physical constraints on
     the Extreme Ultra Violet (EUV) images and photo maps.
Our physical constraints include:
     (i) coronal holes appear darker in EUV images \cite{altschuler1972coronal},
     (ii) they are unipolar in photo map images (positive or negative values only), and
     (iii) they are not allowed to cross magnetic neutral lines (zero crosses in
           photo map images)  \cite{antiochos2007structure}.

\begin{figure}
	\begin{algorithmic}[0]
		\Function{LS}{{\tt EUVI}, {\tt mag\_img}, {\tt init\_img}}
		\State \COMMENT{\textbf{Input:} {\tt EUVI} and {\tt mag} images to process.}
		\State \COMMENT{\textbf{Output:} Segmented image.}\\
		\State $I \gets$ \textbf{smooth} {\tt EUVI} with $15\times 15$ Gaussian kernel
		\State \hspace{1.9cm} with optimization variable $\sigma$.
		\State {\tt g}   $\gets$ $\frac{1}{1+(I_x^2+I_y^2)}$ 
		~\\
		\State\COMMENT{Make {\tt g} zero at magnetic boundaries} 
		\State{\tt p}   \hspace{2mm}$\gets$ \textbf{DetectMagneticCrossLines}({\tt mag\_img})
		\State{\tt pg} $\gets$ (1 - {\tt p} ) .* {\tt g} \\
		
		\State \COMMENT{Init. using combination of Henney-Harvey \cite{Henney2005},}
		\State \COMMENT {SegNet \cite{SegNet}, and FCN \cite{long2015fully}.}
		\State $\phi$ $\gets$ {\tt init\_img} 
		\For{i $\leq$ n }
		\State $\delta$($\phi$) $\gets$ \textbf{Dirac}($\phi$, $\epsilon$) \\
		~\\
		\hspace{0.5 true in}\COMMENT{Use modified edge function {\tt pg}:}
		\State $F_a$ $\gets$ \textbf{areaTerm}($\delta$($\phi$), {\tt pg}) \
		\State $F_e$ $\gets$ \textbf{edgeTerm}($\delta$($\phi$), $\phi$, {\tt pg})
		\State $F_d$ $\gets$ \textbf{Regularize\_distance}($\phi$)\\
		~\\
		\hspace{0.5 true in}\COMMENT{Allow $\alpha$ to vary for optimization:}
		\State $\phi$ $\gets$ $\phi$ + ts$\cdot$($\mu F_d$ + $\lambda F_e$ + $\alpha F_a$)
		\EndFor \\
		\noindent{\bf return} $\delta(\phi)$
		\EndFunction
	\end{algorithmic}
	\caption{
		Level-set segmentation algorithm using
		the modified edge function {\tt pg}.}\label{fig:LevelSets}
\end{figure}

To summarize the level-set segmentation method, we review
     the basic definitions given in \cite{chunming2010}.
Let $p(.)$ be used for defining a regularized distance for
     the level set function ($\phi$),
     $g(.)$ denote the edge function that  should be minimized at image edges,
     $G(.)$ denote the Gaussian, and
     $\nabla$ denote the gradient operator.
     We then define the divergence
     operator using
     $d_p(s) \overset{\vartriangle}{=} p'(s)/s$,
     the edge function
     $g=1/(1+|\nabla G*I |^2) $,
     and a localization function
     $\delta_\epsilon(x)$
     that is zero for  $|x|>\epsilon$ 
     and non-zero for $|x|<\epsilon$.

	We modify the edge function so that it does not allow crossing the magnetic neutral lines.
	This is accomplished by modifying the edge function to be:
	\begin{equation}
	\label{eq:pg}
	pg = (1-p) g
	\end{equation}

where $p$ assumes the value of 1 over the magnetic polarity 
boundaries detected in the
magnetic image and is zero away from the boundary.
Thus, over the magnetic lines, the edge function becomes zero and prevents crossing of the neutral line boundary.

The segmented image is computed by evolving the level set as given by:
\begin{align}
  \frac{\partial \phi}{\partial t} &= \mu {\cal R}_p (\phi) 
                                     + \lambda {\cal L}_{pg} (\phi) 
                                     + \alpha    {\cal A}_{pg} (\phi)
                                     \label{eq:leveldef1}
                                     \intertext{where:}
                                     {\cal R}_p (\phi) &= \rm{div}(\rm{d_p}(|\nabla\phi|)\nabla\phi) 
                                         \,\,\text{is the distance term,} \nonumber \\
  {\cal L}_{pg} (\phi) &=  \delta_\epsilon(\phi)\cdot {\rm div}
                              \left(pg\cdot \frac{\nabla\phi}{|\nabla\phi|} \right)
                           \,\, \text{is the boundary term, and} \nonumber \\
  {\cal A}_{pg} (\phi) &= pg\cdot \delta_\epsilon(\phi) 
                      \quad\text{is an area term.} \nonumber 
\end{align}

We provide a description of the proposed level-set
   segmentation algorithm in 
   Fig. \ref{fig:LevelSets}.  
The approach requires 
joint processing of the EUV and magnetic images.
From \eqref{eq:leveldef1}, we have found that $\alpha$ and
the spatial spread of the Gaussian ($\sigma$) used for computing 
the edge function are the two parameters that can affect overall segmentation
performance.
To find the optimal parameter values, we compare against
the consensus maps, and look for the optimal values:
\begin{equation}
  \min_{\alpha, \, \sigma} \sqrt{[1-{\tt spec}(\alpha, \sigma)]^2 + [1-{\tt sens}(\alpha, \sigma)]^2} 
  \label{eq:basicOpt}
\end{equation}
where ${\tt Spec}$ denotes the (pixel-level) specificity and ${\tt Sens}$ denotes the 
corresponding sensitivity.
The solution of \eqref{eq:basicOpt} gives the optimal values for each image.
For each image, we constrain the optimization problem for 
$\alpha\in[-3, +3], \, \sigma\in [0.2, 1]$.
Over the training set, we select the median values
over the entire set. 

The optimization of \eqref{eq:basicOpt} is challenging since
derivative estimates can be very noisy.
To this end, we use a robust optimization method
based on Pattern-search initialized with $\alpha_0=0, \sigma_0=0.5$.
We refer to \cite{optbook} for details on the optimization procedure.

\begin{figure}[t!]
        \begin{algorithmic}[1]
                \Function{cluster\_matching}{{\tt dates}}\\
                \COMMENT{\textbf{Input:} ~~~{\tt dates} to process. Each date has} 
                \Statex \hspace{1.6cm} an associated list of physical models and a 
                \Statex \hspace{1.6cm} a reference (segmentation) map.\\
                \COMMENT{\textbf{Output:}{\tt ~model\_maps} of the new, missing, and}
                \Statex \hspace{1.6cm} matched coronal holes for each physical model
                \Statex \hspace{1.6cm} and date are stored as separate files.
                \Statex
                \State~\C{Process each date separately}
                \For{ {\tt date} $\in$ {\tt dates}}
                \State \C{Read  and process reference image}
                \State {\tt ref\_map} $\gets$ {\bf load\_ref\_data}({\tt date})
                \State {\tt ref\_map$_{\{+,-\}}$} $\gets$ {\bf pre\_process}({\tt ref\_map})\label{alg:ov_preprocessing1}
                \Statex
                \State~\C{Process associated physical models}
                \For{ {\tt model} $\in \{{\tt model\_1}, \dots, {\tt model\_M}\}$ }
                \State {\tt model\_map} $\gets$ {\bf load\_model}({\tt date}, {\tt model})
                \State {\tt model\_map$_{\{+,-\}}$}  
                \Statex \hspace{3.7cm}$\gets$ {\bf pre\_process}({\tt model\_map})\label{alg:ov_preprocessing2}
                \Statex
                \State~\C{Analyze each polarity separately}
                \For{polarity {\tt p} $\in$ \{ $+$, $-$\}}
                \State {\textbf{\textit{Cluster}}} coronal holes that are are close.
                \State {\textbf{\textit{Detect}}}  coronal hole clusters that are in
                \Statex \hspace{2.9cm} physical maps but not in reference
                \Statex \hspace{2.9cm} map using Mahalanobis distance
                \Statex \hspace{2.9cm}  and store the results in
                \Statex \hspace{2.9cm} {\tt new\_map$_p$} and {\tt missing\_map$_p$}
                \State {\textbf{\textit{Re-cluster}}} remaining coronal holes in 
                \Statex \hspace{2.9cm} {\tt ref\_map} and {\tt model\_map} to
                \Statex \hspace{2.9cm} achieve equal number of clusters.
                \State {\textbf{\textit{Match}}} clusters using \textit{linear programming} 
                \Statex \hspace{2.9cm} and save the results in
                \Statex \hspace{2.9cm} {\tt matched\_map$_p$}                                                            
                \EndFor
                \Statex
                \State~\C{Save maps for each physical model}
                \State {\tt model\_maps} $\gets$  ({\tt new\_map$_p$}, 
                \Statex \hspace{2.2cm} {\tt missing\_map$_p$}, {\tt matched\_map$_p$})
                \Statex \hspace{2.2cm} for polarity {\tt p} $\in$ \{+, -\}.
                \State {\textbf{\textit{Store}}} {\tt model\_maps}
                \EndFor
                \EndFor     
                \EndFunction
        \end{algorithmic}
        \caption{Cluster matching algorithm. The algorithm process the segmentation map
                  (reference map) against the physical models for each given date. 
                  The resulting maps contain the new, missing, and matched coronal holes
                       between the reference map and each physical model.
                  In our example, we have 12 physical maps ($M=12$).     }
        \label{Fig:autoClassOverviewAlg}
      \end{figure} 

      \begin{figure}[t!]
        \begin{algorithmic}[1]  
                \Function{pre\_process}{{\tt map}}\\
                \COMMENT{{\textbf Input:} \hspace{0.2cm} {\tt map} to process.}\\
                \COMMENT{{\textbf Output:}}   {\tt map$_{\{+,-\}}$} contains separate coronal hole 
                \Statex  \hspace{1.4cm} for positively and negatively charged coronal 
                \Statex  \hspace{1.4cm} holes of the same size as the magnetic images,
                \Statex  \hspace{1.4cm} with polar regions removed.
                \Statex
                \State \COMMENT{Extract reference and model maps}
                \State {\tt map} $\gets$ {\small \textbf{extract\_coronal\_hole\_maps}} ({\tt map})
                \State
                \State \COMMENT{Remove small gaps in the maps}
                \State {\tt map} $\gets$ {\bf binary\_close~}({\tt map})
                \Statex
                \State \COMMENT{Resize maps to photomap image size}
                \State {\tt map} $\gets$ {\bf resize~}({\tt ref\_map})
                \Statex
                \State \COMMENT{Separate the positive from the negatively charged}
                \Statex \hspace{1cm} {coronal holes}
                \State {\tt map$_{\{+,-\}}$} $\gets$ {\bf extract\_polarity\_maps~}({\tt map})     
                \Statex
                \State  \COMMENT{Remove regions where there are no observations,}
                \Statex \hspace{0.8cm} and regions with latitude 0 to 30 degrees, and 
                \Statex \hspace{0.8cm} 150 to 180 degrees
                \State {\tt map$_{\{+,-\}}$}  $\gets$  {\bf remove\_regions} ({\tt map$_{\{+,-\}}$})
                \Statex
                \State \Return {\tt map$_{\{+,-\}}$}
                \EndFunction
        \end{algorithmic}  
        \caption{\label{fig:preProcessing}Pre-processing steps for cluster matching.}
\end{figure} 

\subsection{Clustering and Matching}\label{sec:ClusteringAndMatching}
We use the segmented maps as reference maps for comparing
   against the physical models as detailed in Fig. \ref{Fig:autoClassOverviewAlg}.
For each date, we compare each physical model against the reference
   map to determine new, missing, and matching coronal hole clusters. 
In what follows, we provide more details for each step. 
\subsubsection{Pre-processing}\label{sec:ch4preprocessing}
Pre-processing is summarized in Fig. \ref{fig:preProcessing}.
Each physical map is resized from $144\times 172$ to 360$\times$180 using bilinear interpolation
 to match the resolution of the EUV, magnetic, and final segmentation maps.
Then, we remove the polar regions and regions where we have no observations.
We return the positive and negative maps separately for further processing.

\subsubsection{Clustering}\label{sec:ch4clustering_close}
The initial clusters are formed from coronal holes that are close to each other.
To measure distance, let $A$ and $B$ define the set of pixels that belong to 
   two coronal holes of the same polarity.
We define set distance using $d(A, B)=\min_{x\in A, y\in B} d(x, y)$ where
   $d(x,y)$ is based on distance.
Thus, initially, we cluster $A$ and $B$ into the same cluster
   if $d(A, B)<T$ where $T$ is a threshold value that is determined during training.
Here, we note that the relationship is transitive.
In other words, if $d(A, B)<T$ and $d(B, C)<T$, we cluster 
   $A, B, C$ into the same cluster.   

\subsubsection{Detecting new and missed coronal hole clusters}\label{sec:ch4detecting_genrem}
After clustering, there is a need to compare the reference map to the model maps to detect significant differences. We use the Mahalanobis distance to determine coronal hole clusters that are sufficiently close for possible matching.
For computing the Mahalanobis distance, we compare  the physical areas and set distances (minimum physical distances) between coronal hole clusters.  

The remaining coronal hole clusters are assumed to be too far apart to be matchable.
Then, from the remaining ones, coronal hole clusters that are in the reference map but are missing from the model maps are classified as \textit{missing} and are added to the {\tt missing\_map$_p$} (for the $p$-th model map).
Similarly, coronal hole clusters that are in the model map but are missing from the reference map are classified as \textit{new} and are added to the  {\tt new\_map$_p$} (for the $p$-th model map).

\subsubsection{Matching with re-clustering}\label{sec:ch4Matching}
After removing the new and missing coronal hole clusters, the 
remaining ones need to be matched. Unfortunately, we
can still have different numbers of clusters in each map.
Thus, instead of matching  maps having different number
of clusters, we need to first combine them
together to have equal numbers of clusters. Clustering
is accomplished using the minimum physical distance between coronal holes.
Here, we note that our use of the minimum spherical distance takes
   into account wrap-around effects.
The basic idea is to iteratively cluster together all
coronal hole clusters that are separated by a minimal
physical distance until we reach the desired number of
clusters. We introduce linear programming model for
computing an optimal matching between coronal hole clusters.

Let $i$ be used to index clusters in the reference map.
Similarly, let $j$ be used to index clusters in the physical map.
Then, we use  $m_{i,j}$
to denote a possible match between cluster $i$ in the reference map
and cluster $j$ in the physical map.
Thus, $m_{i,j}=1$ when there is a match between the clusters
and $m_{i,j}=0$ otherwise.
We also assign a cost $w_{i,j}$ associated with the matching.
Here, we set the $w_{i,j}$ to be the shortest spherical distance
between the clusters (set distance).
Thus, $w_{i,j}=0$ when the clusters overlap.

Formally, we find an optimal matching by solving:
\begin{align}
 &\min_{m_{i,j}} \,\, \sum_i \sum_j w_{i,j} m_{i, j} 
\label{eq:min_problem} \\
\intertext{subject to:}          
&\sum_i m_{i, j} = 1, \label{eq:constri} \\
&\sum_j m_{i, j} = 1  \label{eq:constrj}\\
&m_{i, j} \in \{0, 1\} 
\end{align}\\
\\
where $m_{i, j}$ denotes the assignment that minimizes
the weighted matching of \eqref{eq:min_problem},
while each cluster can only be assigned to one
other cluster as required by \eqref{eq:constri} and \eqref{eq:constrj}.
This is a typical bipartite matching setup, making matching matrix
created from $m_{i,j}$ to be totally unimodular. As discussed in
\cite{papadimitriou1982combinatorial} this problem when solved with
linear programming will return an integer solution.
\subsection{Classification}\label{sec:Classification}
After cluster matching, we need to make a decision
     on whether the physical model is sufficiently good for forecasting
     applications.
Based on the cluster matching results, we extract the following features for classification:
 (i)   number of new coronal hole clusters,
 (ii)  number of missing coronal holes,
 (iii) total physical area of new coronal hole clusters, 
 (iv)  total physical are of missing coronal hole clusters, and
 (v)   overestimate of physical area of coronal holes as estimated by the physical model.
Here, we note that the area overestimate comes from the need to avoid underestimating
   the impact that the coronal holes will have on the earth.
The basic idea is that the excessive area will likely account for 
   more potential events that should not be missed by the physical models.
        
For the final classifier, we use a Random Forest. Here, we note that a random forest classifier produces a majority classification based on a collection of tree classifiers. For each tree, decisions are represented as inequalities implemented on the five basic features that we have just described. The majority vote can significantly reduce the variance of the resulting classifier (see   
   \cite{Breiman2001,ESLII} for details).
   

\section{Results} \label{sec:results}
The results are summarized in five sections.
First, we provide a description of the dataset in section \ref{sec:ground_truth}.
We then show an example of image segmentation in section \ref{sec:segm}.
We summarize results for coronal hole matching
   in section \ref{sec:detection},
   and final classification
   results in section \ref{sec:classification_results}.

\subsection{Dataset}\label{sec:ground_truth}
The dataset consisted of two Carrington rotations \cite{howard1981surface}
for a total of 50 days. The first Carrington rotation covers the dates 
from 07/13/2010 to 08/09/2010. The second Carrington rotation covers the
dates from 01/20/2011 to 02/16/2011. For each day, we used:
\begin{itemize}
        \item The EUV (synoptic) image (see \cite{altschuler1977high})
        \item Magnetic photo map image
        \item Segmentations by two independent expers
        \item The Consensus map derived from the expert segmentations
        \item Twelve physical model maps manually classified as Good or Bad
               based on their agreement with the corresponding consensus map.
\end{itemize}
For each day we have 12 coronal hole forecasts. 
Hence, for training the random forest classifier, we
consider 600 images that correspond to 12 physical models per day (50 days total).

For level-set segmentation, we used 70\% of the images for training
    and 30\% for testing.    
We report segmentation results on the test set.
Similarly, for physical map classification, we    
   train the classifier on 70\% (419 maps) of the maps and report the 
   results on the remaining 30\% (181 maps).

\subsection{Segmentation}\label{sec:segm}
\begin{figure}[!t]

	\subfloat[EUV image]{ 
		\includegraphics[width=0.47\linewidth]{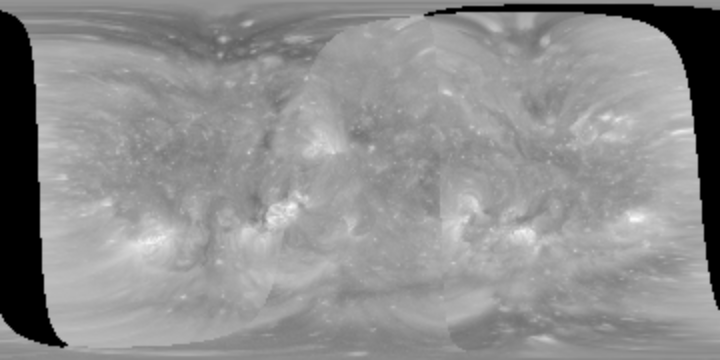}
		\label{subfig:segmentation_EUVImage}}
	~\subfloat[Magnetic photomap image]{ 
		\includegraphics[width=0.47\linewidth]{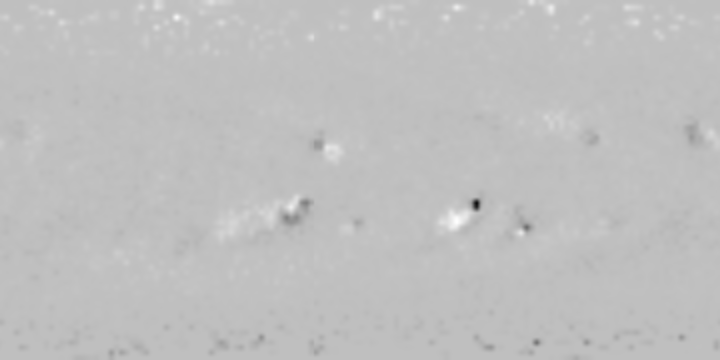}
		\label{subfig:segmentation_magnetic}} \\
	\subfloat[Consensus segmentaion]{ 
		\includegraphics[width=0.47\linewidth]{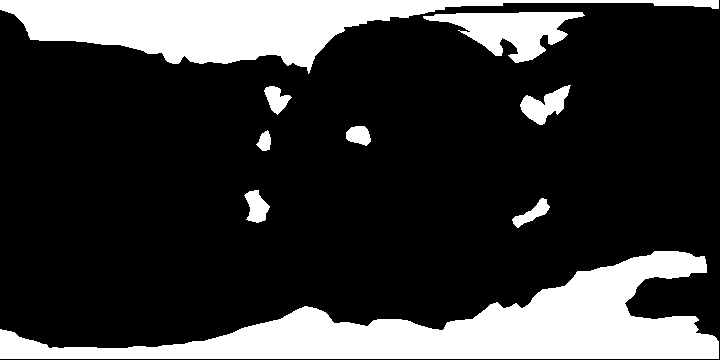}
		\label{subfig:segmentation_consensus}
	}
	\subfloat[Henney-Harvey segmentation with Random Forest selection]{ 
		\includegraphics[width=0.47\linewidth]{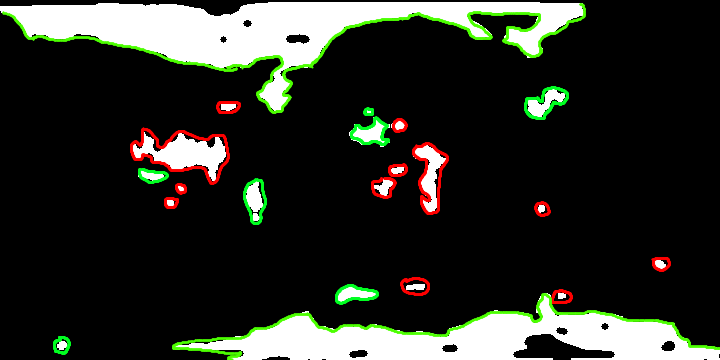}
		\label{subfig:segmentation_hh}
	}\\
	~\subfloat[FCN segmentation with \newline Random Forest selection]{ 
		\includegraphics[width=0.47\linewidth]{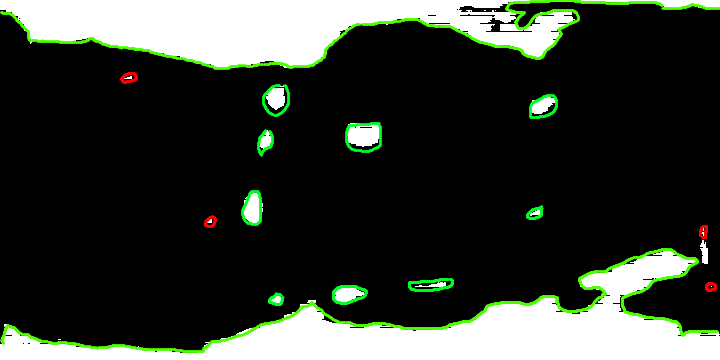}
		\label{subfig:segmentation_fcn}
	}
	~\subfloat[SegNet segmentation with Random Forest selection]{ 
		\includegraphics[width=0.47\linewidth]{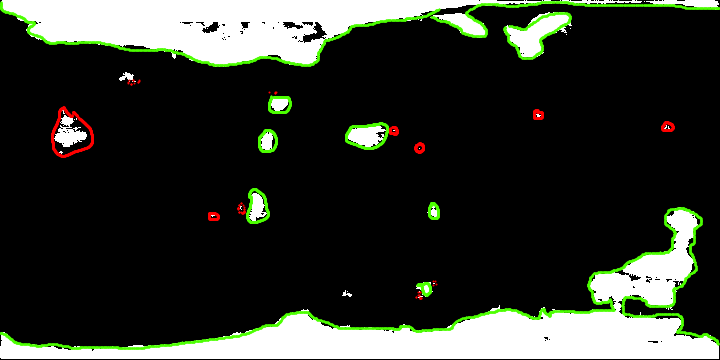}
		\label{subfig:segmentation_segnet}
	}\\
	~\subfloat[Level-sets segmentation \newline overlaid on EUV image]{ 
	\includegraphics[width=0.47\linewidth]{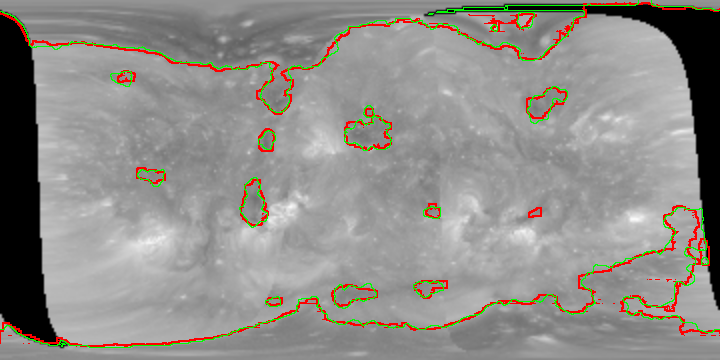}
	\label{subfig:ls}}
	~\subfloat[Proposed segmentation method vs Consensus map]{ 
		\includegraphics[width=0.47\linewidth]{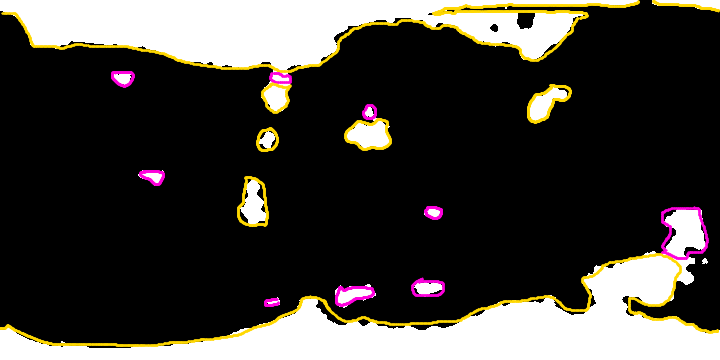}
		\label{subfig:gtls}
	}
	\caption{\label{fig:segmentation}
		Proposed segmentation method for the test date case of 8-7-2010.
	}        
\end{figure}

We demonstrate the performance of the proposed segmentation algorithm 
   in Fig. \ref{fig:segmentation}.
The inputs are the EUV and the magnetic images appearing in 
   (a) and (b).
The ground truth consensus segmentation is shown in (c).
Initial segmentations for
  (d) Henney-Harvey (distance from ({\tt Sens}=1, {\tt Spec}=1) is $d$=0.36), 
  (e) FCN ($d$=0.12), and 
  (f) SegNet ($d$=0.18).
In (d), (e), (f), 
  we use green lines to mark the selected valid coronal holes and
  red to mark coronal holes that were rejected by the random-forest
  coronal hole selector.  
In (g), we show the initial segmentation boundary in red and its final
  evolution in red.
In (h), we compare the final segmentation against the consensus maps ($d$=0.09).
The coronal holes with yellow outlines are in agreement with the Consensus map.
The coronal holes with pink outlines are not in agreement with the Consensus map.
The final segmentation results are also shown on the EUV images in (h).
An unoptimized Matlab implementation of Level-set segmentation takes about 38 seconds
   on a 2.2 GHz Intel Xeon.
For real-time level-set implementations for video data, we refer to \cite{FastLevel}.
For our application, there was no need to improve the speed of the segmentation method.

From the results in Fig. \ref{fig:segmentation}, we can see that
    the Random Classifier selector is very effective. 
A large number of invalid coronal holes were removed
    from the initial segmentations of     
    Figs \ref{fig:segmentation}(d)-(f) (marked red).
On the other hand, we can see that we have also achieved a slight
    over-segmentation in Fig. \ref{fig:segmentation}(h).
As discussed earlier, we want to avoid missing valid
    coronal holes.
As a result, we prefer over-segmentation as opposed
    to under-segmentation.        
As can be seen from the results in Fig. 
    \ref{fig:segmentation}(h), the level-set method propagated
    the coronal-hole boundary to the local magnetic boundary.

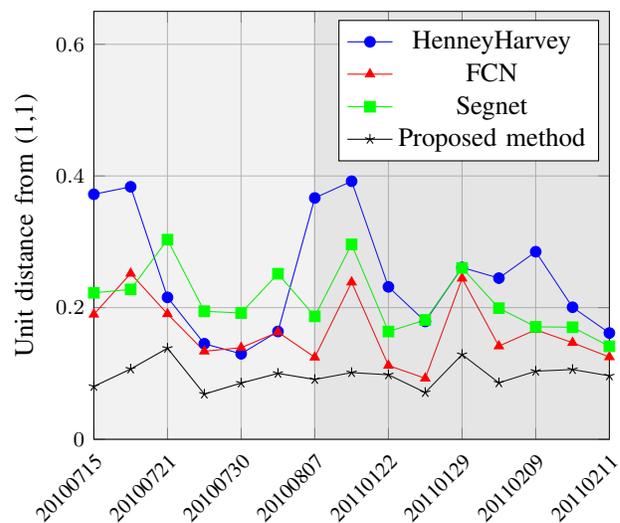
\begin{figure}[!t]
\subfloat{
		\begin{tikzpicture}
\centering
\begin{axis}[
grid=both,
enlarge x limits=false,
symbolic x coords={20100715,
	20100719,
	20100721,
	20100727,
	20100730,
	20100731,
	20100807,
	20110120,
	20110122,
	20110123,
	20110129,
	20110207,
	20110209,
	20110210,
	20110211
},
x tick label style={rotate=45,anchor=north east},
ymin  = 0,
ymax  = 0.65,
ylabel={Unit distance from (1,1)},
]

\addplot [blue, mark=*]table [x=Date, y=Henney_Harvey, col sep=comma] {./plots/segmentation/results.csv};
\addplot [red, mark=triangle*]table [x=Date, y=FCN, col sep=comma] {./plots/segmentation/results.csv};
\addplot [green, mark=square*] table [x=Date, y=Segnet, col sep=comma] {./plots/segmentation/results.csv};
\addplot table [x=Date, y=HH+SegNets+FCN+LevelSets, col sep=comma] {./plots/segmentation/results.csv};
\legend{HenneyHarvey,FCN,Segnet,Proposed method}
\fill[gray,opacity=0.1] ({rel axis cs:0,0}) rectangle ({rel axis cs:0.425,1});
\fill[gray,opacity=0.2] ({rel axis cs:0.425,0}) rectangle ({rel axis cs:1,1});

\end{axis}

\end{tikzpicture}

}
	
\caption{
Comparative segmentation results for different dates.
Results over the first cycle are shown on the left.
Results over the second cycle are shown on the right (shaded region).	
\color{black}
}
\label{plt:segmentation_results}
\end{figure}

To document the accuracy provided by the level-set method,
   refer to Table \ref{tab:SegmentationLevelSets}.
From Table \ref{tab:SegmentationLevelSets}, it is clear
   that level-set segmentations can be used to provide
   substantial improvements over all other methods that
   were considered.
Furthermore, from the results in 
       Table \ref{tab:SegmentationLevelSets},
       it is clear that the addition of the magnetic
       maps did not improve segmentation for FCN and SegNet.
For the proposed method, we considered the full-segmentation
      method described in Fig. \ref{fig:IntroBlkDiag}.
We also show results for individual dates in Fig. \ref{plt:segmentation_results}.
Overall, the proposed method performed extremely well.
In comparison, 
    the error ranges from $2\times$ to $3\times$ for Henney-Harvey (two to three times larger),
    $1.5\times$ to $2\times$ for FCN, 
    $2\times$ to  $3\times$ for SegNet, and
    $3\times$ to  $5\times$ for U-net.

\begin{table*}[!t]
	\centering	
	\caption{
		    Segmentation results as a function of distance from the ideal result
		        ({\tt Sens}=1, {\tt Spec}=1).
		    Level-set segmentation improved all prior methods.    
		    We trained a 3-stage U-net with 7M parameters, a learning rate of $10^{-4}$
		        over 1000 epochs.
		    For the rest of the methods, refer to
		         section \ref{sec:Segmentation} for how the segmentation methods were trained.
		    We list the ratio of the achieved performance over the proposed method in parenthesis.
		    Thus, $1\times$ refers to the performance of the proposed method.
		    A value of $2\times$ implies that the error was twice as large as the proposed method.
		    }
    \label{tab:SegmentationLevelSets}	    
	\begin{tabular}{p{5cm} p{2cm} p{2cm} p{2cm} p{2cm} p{2cm}}
		\toprule
		\textbf{Method} & \multicolumn{5}{c}{\textbf{Distance from} ({\tt sens}, {\tt spec}) = $(1, 1)$}\\
		 & \textbf{min} & $\mathbf{25\%}$ & \textbf{med} & $\mathbf{75\%}$ & \textbf{max}\\
		\midrule
		
		U-Net (EUVI) & 0.29 (4.8$\times$) & 0.31 (3.8$\times$) & 0.34 (3.7$\times$) & 0.36 (3.6$\times$) & 0.39 (3$\times$) \\[0.2cm] 
		
		Henney-Harvey & 0.12 (2.00$\times$) & 0.17 (2.12$\times$) & 0.22 (2.44$\times$) & 0.27 (2.70$\times$) & 0.39(3.00$\times$)\\[0.01cm]
		\textit{Henney-Harvey + Level-Sets} & \textbf{\textit{0.07} (1.16$\times$)} & \textbf{\textit{0.12} (1.50$\times$)} & \textbf{\textit{0.18} (2.00$\times$)} & \textbf{\textit{0.25} (2.50$\times$)} & \textbf{\textit{0.30 (2.30$\times$)}} \\[0.2cm]
		
		FCN (EUVI)& 0.09 (1.50$\times$) & 0.13 (1.62$\times$) & 0.15 (1.66$\times$) & 0.19 (1.90$\times$) & 0.25 (1.92$\times$) \\[0.01cm] 
		FCN (EUVI + Mag) & 0.10 (1.66$\times$) & 0.13 (1.62$\times$) & 0.15 (1.66$\times$) & 0.2 (2.00$\times$) & 0.26 (2.00$\times$) \\[0.01cm] 
		\textit{FCN (EUVI) + Level-Sets} & \textbf{\textit{0.06 (1.00$\times$)} } & \textbf{\textit{0.09 (1.12$\times$)}} & \textbf{\textit{0.11 (1.22$\times$)}} & \textbf{\textit{0.13 (1.30$\times$)}} & \textbf{\textit{0.17 (1.30$\times$)}} \\[0.2cm] 
		
		SegNet (EUVI) & 0.14 (2.33$\times$) & 0.17 (2.12$\times$) & 0.19 (2.11$\times$) & 0.24 (2.40$\times$) & 0.30 (2.30$\times$) \\[0.01cm] 
		SegNet (EUVI + Mag) & 0.17 (2.83$\times$) & 0.23 (2.12$\times$) & 0.36 (2.11$\times$) & 0.44 (2.40$\times$) & 0.58 (2.30$\times$) \\[0.01cm]
		\textit{SegNet (EUVI) + Level-Sets} & \textbf{\textit{0.11 (1.83$\times$)}} & \textbf{\textit{0.13 (1.62$\times$)}} & \textbf{\textit{0.14 (1.55$\times$)}} & \textbf{\textit{0.16 (1.60$\times$)}} & \textit{\textbf{0.21 (1.61$\times$)}}\\[0.2cm]
		
		\textit{Full segmentation method} & \textbf{\textit{0.06 (1$\times$)}} & \textbf{\textit{0.08 (1$\times$)}} & \textbf{\textit{0.09 (1$\times$)}} & \textbf{\textit{0.10 (1$\times$)}} & \textbf{\textit{0.13 
				(1$\times$)}} \\
		\bottomrule
	\end{tabular}
\end{table*}

\begin{table}[!b]
  \caption{Coronal hole matching results. 
  	 \textit{New} refers to coronal holes that appear in the physical model but are not seen
  	     in the consensus segmentation maps.
  	 \textit{Missing} refers to coronal holes that appear in the consensus maps
  	     but are not produced in the physical model.
  	 Note that by definition, a coronal hole cannot be both 
  	      new and missing.        
     The proposed method achieved an overall accuracy of
     92.8\%.}
  \label{tab:matchingConfusionMatrix}
  \centering
  \begin{tabular}{l l l l l }
    \toprule    
    \textbf{Manual}
    &\multicolumn{3}{c}{\textbf{Proposed Method}} & \\
    \cmidrule{2-4}
    & New & Missing & Matched & Total \\    
    \midrule
    New & \textbf{1958} & 0 & 22 & 1980\\    
    Missing & 0 & \textbf{868} & 3 & 871\\
    Matched & 416 & 360 & \textbf{7522} & 8298\\
    \midrule
    Total & 2374 & 1228 & 7547 & 11149\\
    \bottomrule
  \end{tabular}
\end{table}

\begin{figure}[!b]
	\subfloat[Clustered, positive polarity consensus map.]
	{ \includegraphics[width=0.47\linewidth]{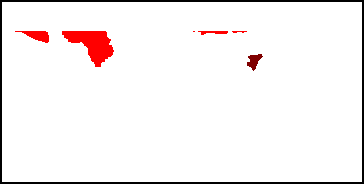}}
	~~\subfloat[Clustered, positive polarity model map.]
	{ \includegraphics[width=0.47\linewidth]{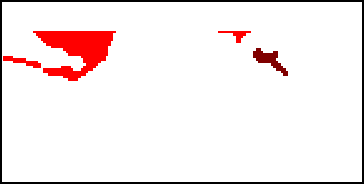}}
	
	\subfloat[Matching clusters from Consensus (left) and model (right) for positive polarity.]
	{ \includegraphics[width=0.94\linewidth]{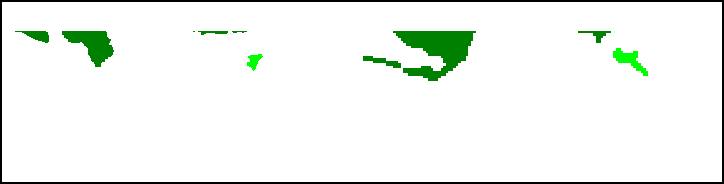}}
	
	\subfloat[Clustered, negative polarity consensus map.]
	{ \includegraphics[width=0.47\linewidth]{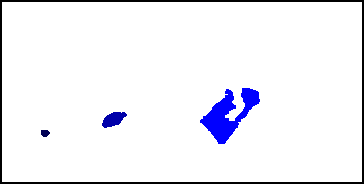}}
	~~\subfloat[Clustered, negative polarity model map.]
	{ \includegraphics[width=0.47\linewidth]{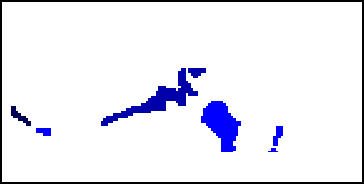}}        
	
	\subfloat[Matching clusters from consensus (left) and model (right) for negative polarity.]
	{ \includegraphics[width=0.94\linewidth]{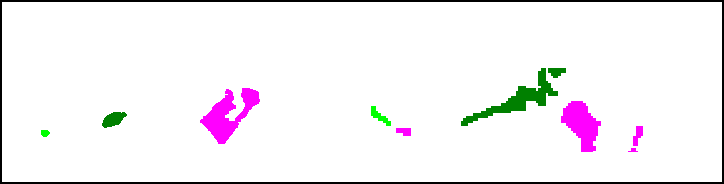}}
	
	\caption{Coronal hole cluster matching example (02-04-2011). Matched clusters share the same color.}
	\label{fig:ch5Matching_good}
\end{figure}

\subsection{Coronal hole matching}\label{sec:detection}
We present the results from coronal hole 
     matching between the Physical maps and the Consensus maps
     in Table \ref{tab:matchingConfusionMatrix}.
The results are compared against manual matching each coronal hole.
Overall, at $92.8\%$, the method produced 7,522 matches,
   in agreement with manual labeling.
The proposed approach failed to match 776 coronal holes
    (776=416+360) while it overmatched 25 (25=22+3) of them.
Thus, compared to the human expert,
    the proposed approach tends to undermatch.       
Nevertheless, at $92.8\%$, the proposed method performed
    very well.

\begin{figure}
        \subfloat[Clustered, positive polarity consensus map.]
        { \includegraphics[width=0.47\linewidth]{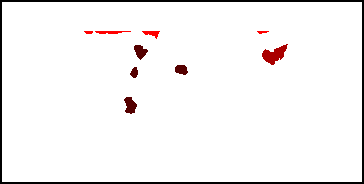}
                \label{subfig:ch5consensus_cluster}}
        ~~\subfloat[Clustered, positive polarity  model map.]
        { \includegraphics[width=0.47\linewidth]{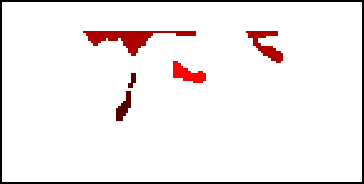}}        
        
        \subfloat[Matching clusters from consensus (left) and model (right) for positive polarity.]
        {\includegraphics[width=0.94\linewidth]{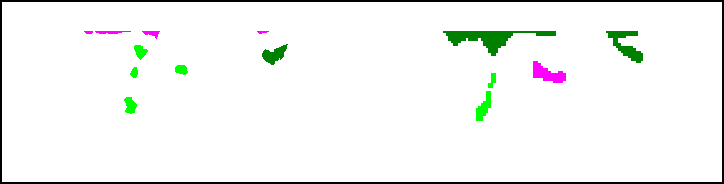}}
        
        \subfloat[Clustered, negative polarity consensus map.]
        { \includegraphics[width=0.47\linewidth]{figures/matching_example/20100807_1ref_pos_matchable_clus.png}}
        ~~\subfloat[Clustered, negative polarity model map.]
        { \includegraphics[width=0.47\linewidth]{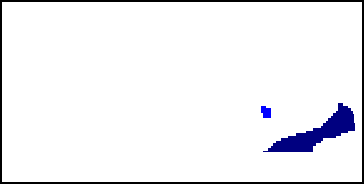}}        
        
        \subfloat[Matching clusters from consensus (left) and model (right) for negative polarity.]
        {\includegraphics[width=0.94\linewidth]{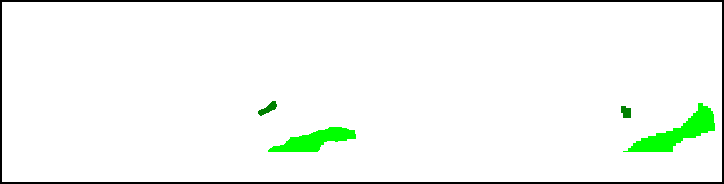}}
        
        \caption{Difficult example of coronal hole matching showing issues in the algorithm 
                (07-08-2010). Matched clusters share the same color. }        
        \label{fig:ch5Matching_bad}
\end{figure}

Results of matching using linear programming are demonstrated in Figs. \ref{fig:ch5Matching_good} 
and \ref{fig:ch5Matching_bad}.
Fig. \ref{fig:ch5Matching_good} shows an example where automated matching
agrees with manual matching.
Fig. \ref{fig:ch5Matching_bad} shows an example where there are significant
differences between automated and manual matching (see pink cluster).

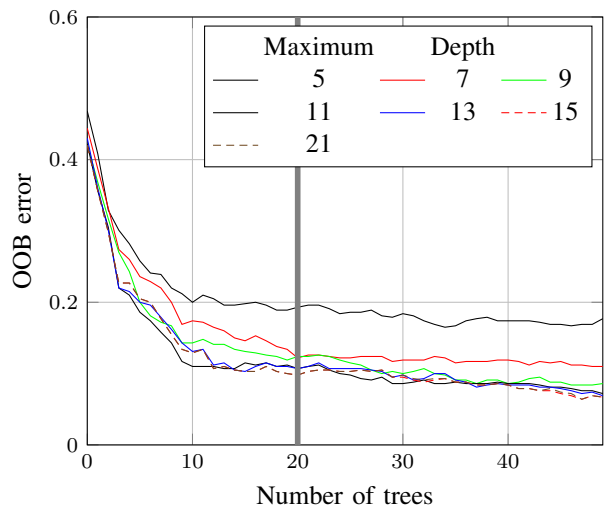
\begin{figure}[!t]
		\begin{tikzpicture}
		\centering
		\begin{axis}[
		grid=both,
		enlarge x limits=false,
		xlabel={Number of trees},
		ymin  = 0,
		ymax  = 0.6,
		ylabel={OOB error},
		legend style ={no markers,at={(0.98,0.98)},},
		legend columns=3
		]

        \addlegendimage{empty legend}
		\addlegendentry{Maximum}		
		\addlegendimage{empty legend}
		\addlegendentry{Depth}
        \addlegendimage{empty legend}
		\addlegendentry{}
		\addplot[mark=none] table [x=trees, y=five, col sep=comma] {./plots/random_forest_max_depth_and_num_trees/data.csv};
		\addplot table [x=trees, y=seven, col sep=comma] {./plots/random_forest_max_depth_and_num_trees/data.csv};
		\addplot[green] table [x=trees, y=nine, col sep=comma] {./plots/random_forest_max_depth_and_num_trees/data.csv};
		\addplot table [x=trees, y=eleven, col sep=comma] {./plots/random_forest_max_depth_and_num_trees/data.csv};
		\addplot table [x=trees, y=thirteen, col sep=comma] {./plots/random_forest_max_depth_and_num_trees/data.csv};
		\addplot table [x=trees, y=fifteen, col sep=comma] {./plots/random_forest_max_depth_and_num_trees/data.csv};
		\addplot table [x=trees, y=twentyone, col sep=comma] {./plots/random_forest_max_depth_and_num_trees/data.csv};
		\draw[line width=0.75mm, gray] (20,0) -- (20,0.6);

		\addlegendentry{5}
		\addlegendentry{7}
		\addlegendentry{9}
		\addlegendentry{11}
		\addlegendentry{13}
		\addlegendentry{15}
		\addlegendentry{21}
		\end{axis}
		\end{tikzpicture}
\caption{
		Out-of-bag (OOB) error used to determine the optimal parameters for the Random Forest.}
\label{plt:random_forest_n_trees_and_max_depth}
\end{figure}

\subsection{Classification results}\label{sec:classification_results}
We present the results from training the Random Forest classifier in Fig. \ref{plt:random_forest_n_trees_and_max_depth}.
To train the classifier, we further split the training set into 
   a 70\% training set and a 30\% test set as recommended in
   \cite{ESLII}.
The results over the test set (part of the original training set) 
   were used for determining the optimal parameters.
The out-of-bag (OOB)  error is shown as function of the number of trees and the maximum tree depth
   in Fig. \ref{plt:random_forest_n_trees_and_max_depth}.
From the graph, we determine that the optimal configuration will be to use
   20 decision trees at a maximum depth of 11.

We present classification results over the training set
   in Table \ref{tab:ch5classificaiton_results_auto}.

Table \ref{tab:ch5classificaiton_results_auto} summarizes the results over
   the remaining 30\% of the data that were left for testing.
Overall, the classifier had an accuracy of 95.5\%.

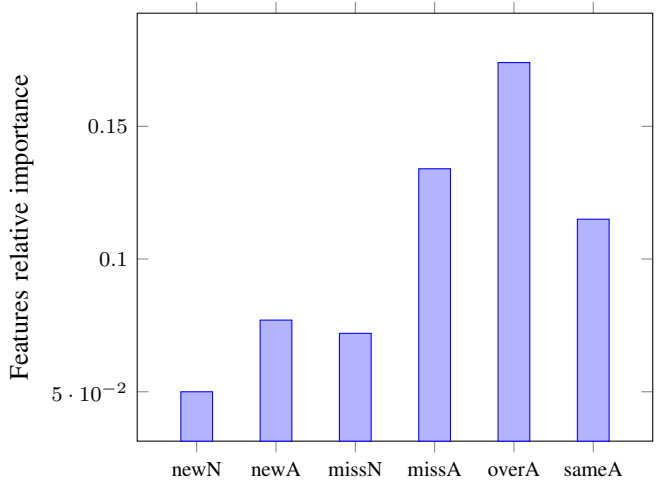
\begin{figure}[!t]
	\begin{tikzpicture}
		\begin{axis}[
		ylabel=Features relative importance,
		enlargelimits=0.15,
		legend style={at={(0.5,-0.15)},
			anchor=north,legend columns=-1},
		ybar,
		bar width=12pt,
		 symbolic x coords={newN,newA,missN,missA,overA,sameA}, 
		 xtick=data,
		]
		\addplot 
		coordinates {(newN,0.05) (newA,0.077)
			(missN,0.072) (missA, 0.134) (overA,0.174)
		    (sameA,0.115)};
		
		\end{axis}
	\end{tikzpicture}
	        \caption{
	        	Relative importance of features from random forest. The features from left to right are 
		\texttt{newN}=number of new coronal holes, 
		\texttt{newA}=image area of new coronal holes,
		\texttt{missN}=number of missing coronal holes,
		\texttt{missA}=image area of missing coronal holes,
		\texttt{overA}=spherical area overestimated by model,
		\texttt{sameA}=spherical area overlap.}
	\label{fig:randomForest_features}
\end{figure}

We also examine the relative importance of each feature in the Random Forest
   classifier and report the results in Fig. \ref{fig:randomForest_features}.
Here, we note that the relative importance is a measure of
    the predictive performance of each variable   \cite{ESLII}.
From Fig. \ref{fig:randomForest_features}, the 
    most important features come from the area features.
Most importantly, a bad physical map will be the result of large amounts
    of missing area.
Similarly, a good physical map will share the same area or have slightly more
    area.             

\begin{table}[!b]
	\caption{
		Classification results for selecting physical maps for forecasting.
		We classify a physical map as \textit{good} if it should be used for forecasting.
		On the other hand, a \textit{bad} physical map is not appropriate for forecasting.
		Overall classification accuracy is $95.5\%$.
	}
	\label{tab:ch5classificaiton_results_auto}
	\centering
	\begin{tabular}{l  p{1cm} p{0.5cm} l}
		\toprule
		\textbf{Manual}	
		&\multicolumn{2}{c}{\textbf{Proposed Method}}
		& \\
		\cmidrule{2-3}	
		& Bad & Good & Total \\
		\midrule         
		Bad      & \textbf{76}  & 3           & 79  \\
		Good     & 5            & \textbf{97} & 102 \\
		\midrule
		Total    & 81  & 100 & 181\\
		\bottomrule
	\end{tabular}
\end{table}

\section{Conclusion}\label{sec:conclusion}

The manuscript summarizes the development of 
  new solar image analysis methods to support:
  (i)    accurate segmentation of coronal holes, 
  (ii)   reproducible classification of physical models based on a manual
         protocol, 
  (iii)  coronal hole clustering and matching, and
  (iv)   a fully automated method for physical model classification. 
The performance of each method is validated   
    against human experts. 

The proposed segmentation method significantly outperformed 
    several other segmentation methods.
There are two fundamental reasons that explain the
    substantially improved segmentation performance.
First, the accuracy of auto-encoder based segmentation methods is
    fundamentally limited by the use of downsampling
    and upsampling operations as documented in \cite{shelhamer2016fully}
    and boundary uncertainty as documented in \cite{BayesianSegNet}.
Second, as for all deep learning methods, the need
    to learn a substantial number of weights requires
    the use of large datasets.
More generally,
    for accurate segmentation,
    we propose the use of level-set methods
    that are initialized by deep-learning methods.
The increased accuracy comes from
    the fine-tuning of the segmentation boundary
    by the level-set method.
On the other hand, initialization by deep learning
    methods enables level-set methods         
    to evolve from a boundary that
    is closer to the optimal one.

The proposed method supports 
    forecasting based on space weather models.
Nevertheless, prior to deployment, 
    the approach needs to be validated using 
    large-scale studies.
The code and datasets are available at  
    \url{https://github.com/venkatesh369/HeliosTransactions}.

\section* {Acknowledgments}
The authors would like to acknowledge Dr. R. Hock and Callie Darsey
for contributing to the manual protocols and helping with the manual segmentation.
We would also like to acknowledge the support of
   Dr. C. Henney who provided his segmentation method.


\begin{IEEEbiography}[{\includegraphics[width=1in,height=1.25in,clip,keepaspectratio]{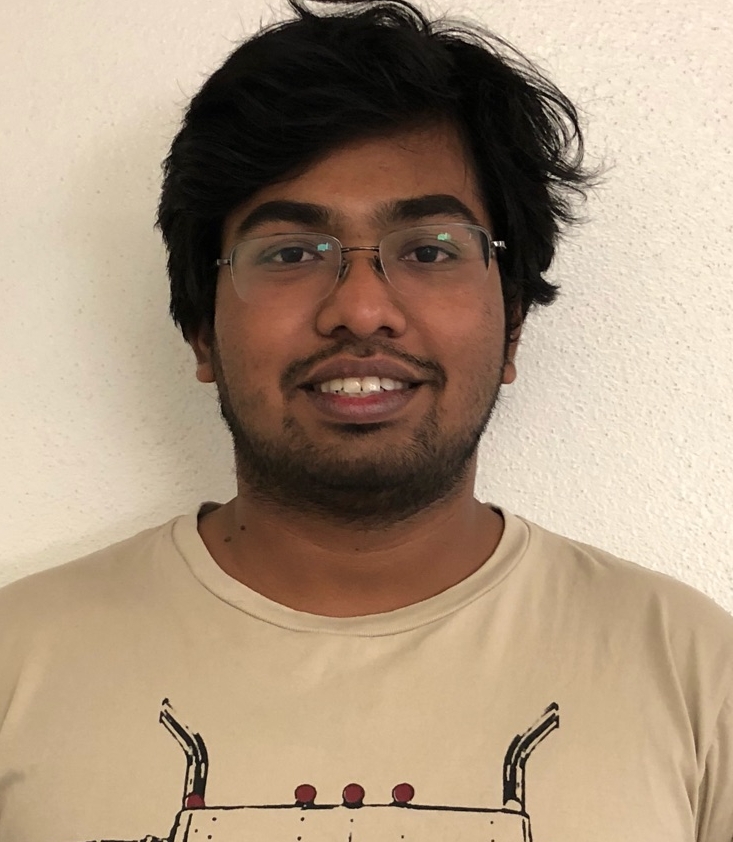}}]{Venkatesh Jatla}
        Venkatesh Jatla received his B.Tech. degree in electrical and communications engineering (ECE) in 2011 from
        VIT, Tamilnadu, India and the M.S. degree in computer engineering in 2016 from University of New Mexico, Albuquerque, USA. He is currently working on his Ph.D. in computer engineering with Dr. Marios Pattichis. His current research
        interests include video analysis, video compression, image processing and machine learning. He is also interested
        in real time video delivery over http and webRTC.
\end{IEEEbiography}
\begin{IEEEbiography}[{\includegraphics[width=1in,height=1.25in,clip,keepaspectratio]{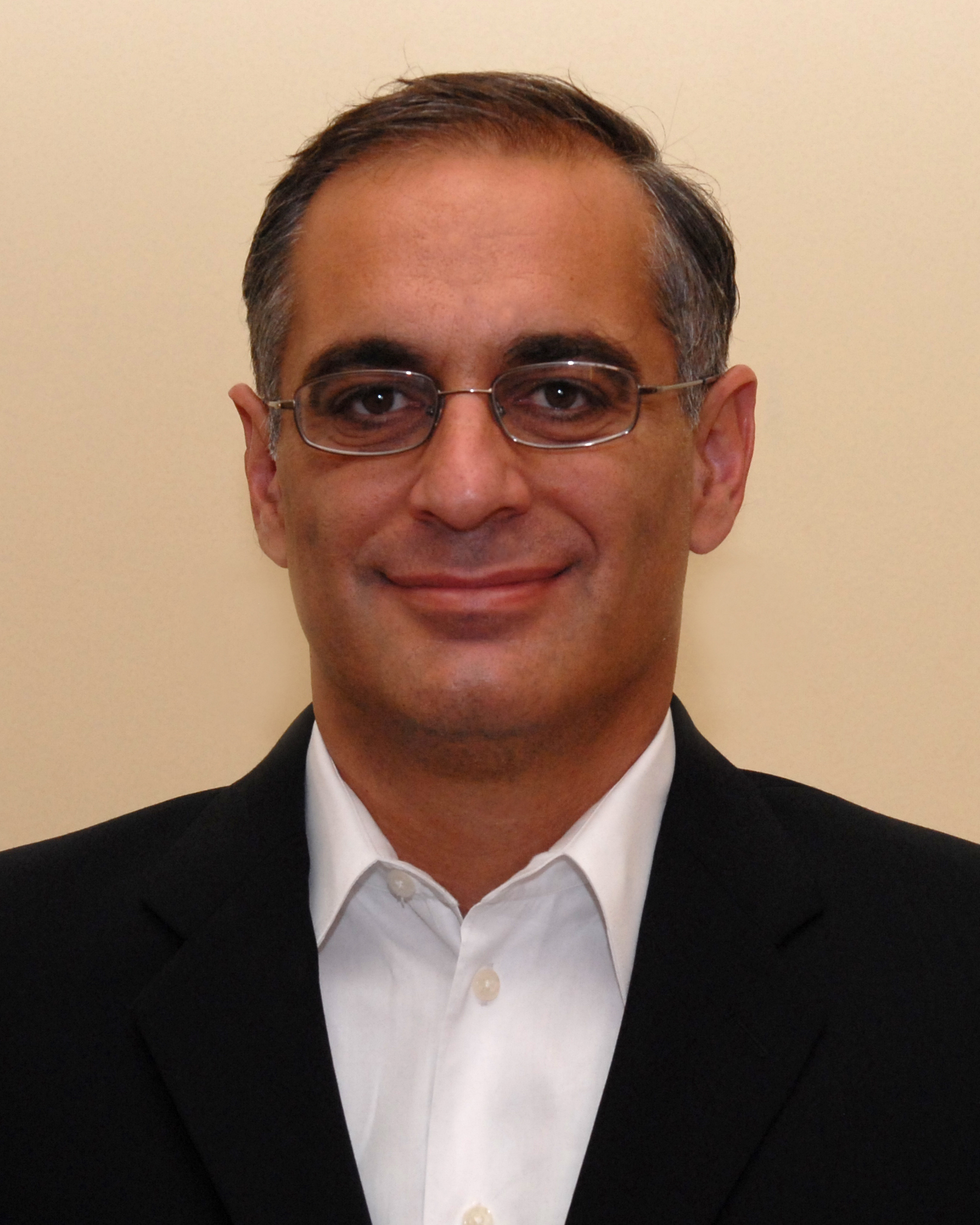}}]{Marios Pattichis}
Marios  Pattichis  (M’99,  SM’06) received  the  B.Sc. (High  Hons.  and  Special  Hons.)
   degree  in  computer  sciences  and  the  B.A.  (High Hons.)  degree  in  mathematics,  
   both  in  1991,  the M.S.  degree  in  electrical  engineering  in  1993,  and
   the  Ph.D.  degree  in  computer  engineering  in  1998,
   all  from  the  University  of  Texas,  Austin.  
He  is currently  a  Professor and Associate Chair  
   with  the  Department  of  Electrical and Computer Engineering, 
   at the University of New Mexico  (UNM) in  Albuquerque.  
His  current  research interests  include  digital  image and video  
   processing, video communications,   
   dynamically  reconfigurable   hardware  architectures,  
   and biomedical and space image-processing applications.
Dr.  Pattichis  is  currently  
   a senior asscoiate editor with the IEEE Transactions on Image Processing.
He has served as a senior  associate  editor  of  the
   IEEE Signal Processing Letters, 
   an associate editor for the IEEE Trans-actions on Image Processing, 
   IEEE Transactions on Industrial Informatics,
   and as a guest associate editor for the 
   IEEE Transactions on Information Technology in Biomedicine. 
He was the general chair of the 2008 IEEE Southwest Symposium on Image Analysis and 
   Interpretation.  
He  was a  recipient  of  the  2004  Electrical  and  Computer  Engineering  Distinguished
   Teaching Award at UNM. For his development of the digital logic design labs
   at  UNM  he  was  recognized  by  the  Xilinx  Corporation  in  2003  and  by  the
   UNM School of Engineering’s Harrison faculty excellent award in 2006. 
He was a founding Co-PI of the COSMIAC research center at UNM. 
At UNM, he is currently the director of the image and video Processing and Communications Lab (ivPCL)
\end{IEEEbiography}
\begin{IEEEbiography}[{\includegraphics[width=1in,height=1.25in,clip,keepaspectratio]{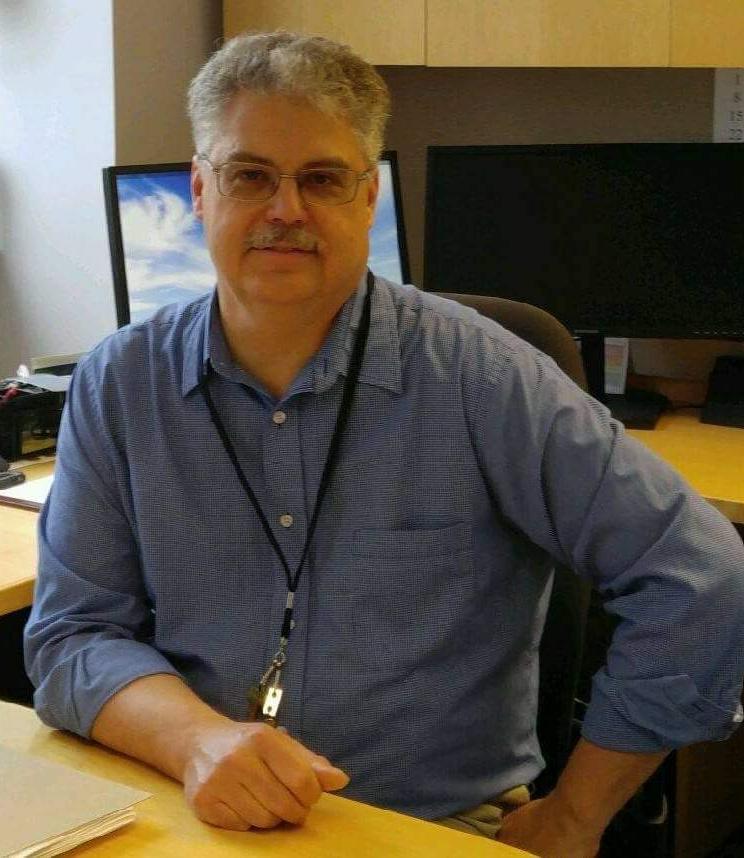}}]{Charles Nick Arge}
C. Nick Arge received his B.S. in Physics at the University of Arizona in 1985, an M.S. degree in Physics at the University of Minnesota in 1988, and a Ph.D. in Physics in 1997 at the University of Delaware. He worked for the University of Colorado \& NOAA/Space Environment Center (now Space Weather Prediction Center) from 1996-2003. From 2004-2016 he worked at the Air Force Research Laboratory (AFRL), Space Vehicles Directorate. In late 2016 he moved to NASA Goddard Space Flight Center where he currently serves as the Chief of the Solar Physics Laboratory, which is comprised of more 100 staff and affiliates. Dr. Arge is also an Adjunct Professor in the Department of Physics and Astronomy at the University of New Mexico and in Department of Astronomy at New Mexico State University. He is a member of the American Astronomical Society (AAS), the AAS Solar Physics Division, and the American Geophysical Union. Dr. Arge received the NOAA Research's Outstanding Scientific Paper Award in 2004. He received AFRL Awards for Technology Transfer Achievement in 2006 \& 2008, Senior Leadership in 2010 \& 2014, and Team Publication in 2015. He received an AFRL Outstanding Mentor Award in 2014 and a Federal Executive Board Public Service Award: Professional, Administrative, Technical in 2015. He was the lead of three Air Force Office of Scientific Research Star Teams. Dr. Arge does basic and applied research in the areas of coronal and solar wind modeling. He is co-developer of the Wang-Sheeley-Arge (WSA) coronal and solar wind model, which currently runs operationally at the NOAA National Centers for Environment Prediction (NCEP).
\end{IEEEbiography}

\end{document}